%% file: acl_latex.tex
\documentclass[11pt]{article}

\usepackage[preprint]{acl}

\usepackage{times}
\usepackage{latexsym}
\usepackage{amssymb}
\newcommand{\emailgrp}[2]{\texttt{\textbraceleft #1\textbraceright @#2}}
\usepackage[T1]{fontenc}

\usepackage[utf8]{inputenc}

\usepackage{microtype}

\usepackage{inconsolata}

\usepackage{graphicx}
\usepackage{booktabs}
\usepackage{booktabs}
\usepackage{array}
\usepackage{multirow}
\usepackage{adjustbox}  
\usepackage{caption}    
\usepackage{makecell}
\usepackage{listings}
\usepackage{xcolor}
\usepackage{pifont} 
\usepackage{algorithm}
\usepackage{algorithmic}
\usepackage{tabularx}
\usepackage{subcaption}
\usepackage{float}
\usepackage{amsmath}
\usepackage[utf8]{inputenc}
\usepackage[T1]{fontenc}
\newcolumntype{Y}{>{\centering\arraybackslash}X}
\definecolor{codegreen}{rgb}{0,0.6,0}
\definecolor{codegray}{rgb}{0.5,0.5,0.5}
\definecolor{codepurple}{rgb}{0.58,0,0.82}
\definecolor{codeblue}{rgb}{0.0, 0.0, 0.8} 
\definecolor{backcolour}{rgb}{0.98,0.98,0.98}
\newcommand{\method}{\textsc{NGDBench}}
\lstdefinestyle{cypherstyle}{
  breaklines=true,
  breakatwhitespace=false,
  columns=fixed,
  keepspaces=true,
  showstringspaces=false,
  basicstyle=\ttfamily\small\linespread{1.12}\selectfont,
  aboveskip=6pt,
  belowskip=6pt,
  lineskip=1pt,
  xleftmargin=1.5em,
  frame=single,
  framesep=4pt
}

\title{NGDBench: Towards Neural Graph Data Management}

\author{%
  \textbf{Yufei Li*}$^{1}$,
  \textbf{Yisen Gao*}$^{1}$,
  \textbf{Jiaxuan Xiong}$^{2}$,
  \textbf{Jiaxin Bai}$^{3}$\\[0.2em]
  \textbf{Shijie Zhong}$^{4}$,
  \textbf{Haoyu Huang}$^{1}$,
  \textbf{Zhongwei Xie}$^{1}$,
  \textbf{Hongting Tsang}$^{1}$,
  \textbf{Yangqiu Song}$^{1}$
  \\[0.25em]
  \normalfont
  $^{1}$HKUST \quad
  $^{2}$Beijing Institute of Technology \\
  $^{3}$Hong Kong Baptist University \quad
  $^{4}$Guangdong University of Technology
  \\[0.25em]
  \emailgrp{ylivm, ygaodi, hhuangcp, zxiebk, httsangaj, yqsong, zwanggc}{connect.ust.hk} \\
  \texttt{xiongjiaxuan@bit.edu.cn} \quad
  \texttt{baijiaxin@hkbu.edu.hk} \quad
  \texttt{zhongshijie2@mails.gdut.edu.cn}
}

\begin{document}
\maketitle
\begin{abstract}
Data critical to real-world decision-making is increasingly found within organizations. Such data is heterogeneous, constantly evolving, and only imperfectly captured. However, current data management systems remain largely passive, retrieving what is explicitly stored while offering limited support for uncovering implicit structure or reasoning under noise, incompleteness, and continuous updates. We argue that next-generation data management requires neural capabilities, which can uncover complex latent relationships, distinguish reliable signals from noise, and remain consistent as the underlying data state evolves. To support this direction, we introduce \method{}, a benchmark across five domains that unifies structured and unstructured sources. \method{} adopts a graph view because graphs provide a flexible abstraction for modeling complex systems, capturing latent relationships, and subsuming structured formats such as relational tables. Each instance pairs a clean latent graph with a realistically perturbed observed graph. \method{} supports full Cypher queries and dynamic data management operations. Evaluations of state-of-the-art Text-to-Cypher by LLMs and GraphRAG pipelines reveal that current neural query methods remain sensitive to noise and struggle with dynamic state tracking, highlighting the need for resilient, inference-capable data management. Our code is available at https://github.com/HKUST-KnowComp/NGDBench.
\end{abstract}

\input{sec/intro}

\input{sec/related_work}
\input{sec/method}

\input{sec/experiment}
\input{sec/conclusion}
\newpage
\bibliography{acl_latex}

\appendix
\input{sec/appendix}

This is an appendix.

\end{document}

%% file: sec/intro.tex
\begin{table*}[t]
\centering
\caption{Comparison of Recent Neural Graph Reasoning Benchmarks. Existing datasets focus on static reasoning with limited logic (FOL/EPFO), whereas our NGDB-Dataset supports full database capabilities including Cypher queries, aggregations, and dynamic updates. }
\label{tab:comparison}
\resizebox{\textwidth}{!}{%
\begin{tabular}{lllcccc}
\toprule
\textbf{Benchmark / Dataset} & \textbf{Data Format} & \textbf{Operators} & \textbf{Multi-variable} & \textbf{Aggregation} & \textbf{Dynamic} & \textbf{General Graph Query} \\
& & & \textbf{Query} & \textbf{Support} & \textbf{Updates} & \textbf{(e.g., Cypher)} \\
\midrule
Q2B~\cite{RenHL20} & Id-based Triples & EPFO (AND, OR, $\exists$) & No & No & No & No \\
LitCQD~\cite{DemirWLNH23}& Id-based Triples \&  Numbers & EPFO + Numeric Operators & No & Yes & No & No \\
NRN~\cite{BaiLLYYS23}& Id-based Triples \&  Numbers & FOL + Numeric Operators & No & No & No & No \\
$\text{EFO}_{1}$-CQA~\cite{WangYS21} & Id-based Triples & FOL (AND, OR, NOT $\exists$) & No & No & No & No \\
$\text{EFO}_{k}$-CQA~\cite{YinWFS23} & Id-based Triples & FOL (AND, OR, NOT $\exists$) & Yes & No & No & No \\
\midrule
\textbf{\method{} (Ours)} & \textbf{Ids, Numbers, and Strings} & \textbf{Cypher Operators} & \textbf{Yes} & \textbf{Yes} & \textbf{Yes} & \textbf{Yes} \\
\bottomrule
\end{tabular}%
}
\end{table*}
\section{Introduction}

\begin{figure}[h]
    \centering
    \includegraphics[width=0.47\textwidth]{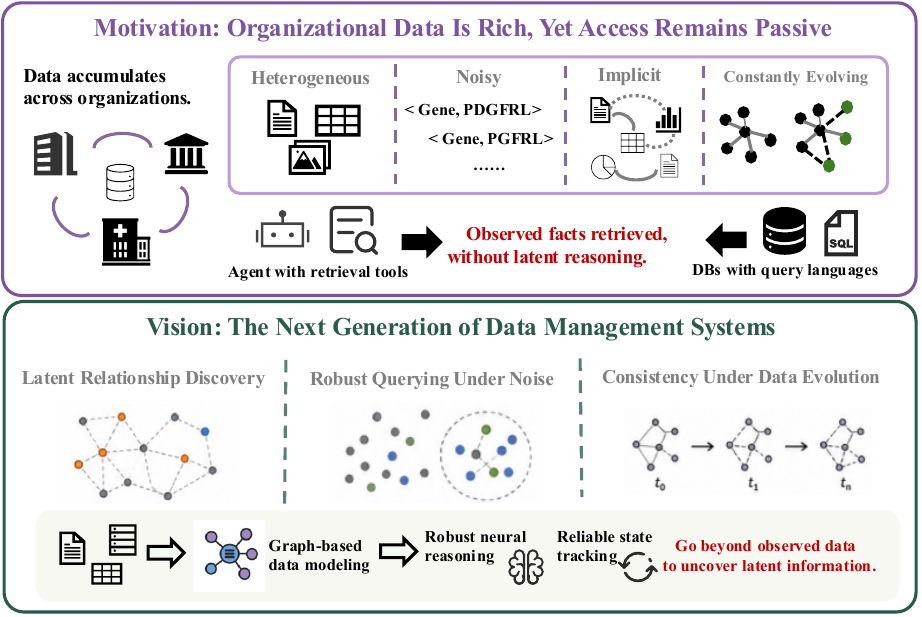}
    \caption{Motivation and Vision: Overcoming the limitations of passive data access through robust graph-based neural data reasoning.}
    \label{fig:motivation}
\end{figure}

As high-quality internet data become increasingly scarce\cite{10.5555/3692070.3694094}, critical decision-making data increasingly reside within organizations, such as enterprises and financial institutions\cite{kayali2025mind}. These repositories, spanning relational, graph, and enterprise systems, are heterogeneous, continuously evolving, often noisy and with implicit information, like hidden entities and latent relation which are lost in the process of recording \cite{abaskohi2026drbenchrealisticbenchmarkenterprise}. There are two dominant ways to access these data: Traditional databases support precise symbolic queries but rely on explicitly defined schemas and stored facts. Vector databases retrieve semantically similar items in embedding space, but they are not designed to perform deeper data analysis beyond similarity matching, such as complex logical queries and reasoning over implicit relations. AI agents access data through interfaces such as RAG, Text-to-SQL\cite{yu-etal-2018-spider}, or Text-to-Cypher\cite{ozsoy2024text2cypherbridgingnaturallanguage}, which similarly expose observed facts rather than latent implications. As a result, the current data stack remains largely passive: it retrieves what is stored, but rarely infers what is implied.

Consider an AI agent performing financial risk analysis on a graph database. Querying only observed records via Text-to-Cypher over noisy or incomplete data fails to uncover latent fraud rings and hidden relations. This gap between observed data and the real-world state fundamentally hinders reliable AI understanding to the real environment.

We therefore argue for a new form of data management system behind AI applications: a \emph{neural database}. Unlike conventional databases that rely on symbolic matching or similarity search over stored records, a neural database should combine precise query execution ablity with neural inference—uncovering latent relationships, distinguishing reliable signals from noise and reliably tracking data state as data evolves.

Graph-structured data naturally supports this vision. Graphs flexibly capture entities, attributes, relations, and multi-hop dependencies~\cite{reddit_graph_db,neo4j_why_graph}, expose latent structure from unstructured sources~\cite{lai2025graphyourdataendtoendmodeling}, and can subsume relational tables~\cite{kim2024cartepretrainingtransfertabular}. As latent-world reasoning is fundamentally relational, graphs offer a suitable substrate for combining symbolic data management with neural inference.
Existing systems move only partway toward this goal. Neural Graph Database connects graph databases with GNN-based prediction \cite{pmlr-v198-besta22a}, while neural complex query answering methods reason over incomplete graphs through logical query embeddings \cite{ren2023neuralgraphreasoningcomplex}. Yet they are mostly restricted to Existential First-Order logic rather than full database queries \cite{WangYS21,YinWFS23}, assume the observed graph as the main source of truth despite invalid observed facts and missing valid facts \cite{huang2025information,drummond2006open}, and provide limited support for state-changing operations such as insertion, deletion, and update. Meanwhile, existing mainstream graph datasets \cite{hu2020open} are often homogeneous, modality-specific, or static. As shown in Table~\ref{tab:comparison}, they do not sufficiently test whether systems can execute full database-style queries over heterogeneous, noisy, incomplete, and dynamically evolving graph data.
 
To support further development in this direction, we introduce \method{}, a benchmark for graph-based neural data management. \method{} spans five domains, including financial, social, biomedical, AI-agent tool-use, and enterprise reporting scenarios, integrating structured records with unstructured sources under a unified Labeled Property Graph schema. Each instance pairs a clean \emph{latent} graph with a realistically perturbed \emph{observed} graph, covering structural and attribute-level imperfections such as missing edges, spurious connections, schema inconsistencies, and corrupted entity descriptions.

Unlike prior benchmarks focused on elementary logical operations, \method{} supports the \textbf{full Cypher query language}, covering complex pattern matching, variable-length paths, numerical aggregations, and dynamic operations such as \texttt{CREATE}, \texttt{UPDATE}, and \texttt{DELETE}. We evaluate \method{} on two widely used neural query paradigms, Text-to-Cypher and LLM-based GraphRAG, which perform graph reasoning by either translating language into symbolic queries or retrieving graph-grounded evidence. Although more promising than purely symbolic matching for handling noise and latent associations, they remain vulnerable to noisy observations and struggle with analytical reasoning and dynamic state tracking.



Our contributions are summarized as follows: 
(1) \textbf{Neural data management vision.} We argue for moving beyond passive record retrieval toward neural inference over latent graph structure, enabling robust querying under noise and consistency under updates. 
(2) \textbf{Neural graph data management benchmark.} We introduce \method{}, a five-domain benchmark that unifies structured and unstructured sources, pairs clean latent graphs with perturbed observations, and supports full Cypher queries with dynamic operations. 
(3) \textbf{Evaluation of current paradigms.} We evaluate state-of-the-art Text-to-Cypher and GraphRAG methods, revealing limitations in noise robustness, analytical reasoning, and dynamic state tracking.

%% file: sec/related_work.tex

\section{Related Work}
\paragraph{Neural Graph Database.} Existing definitions of Neural Graph Databases (NGDBs) generally follow two paradigms. One extends traditional graph databases by preserving symbolic storage, such as Labeled Property Graphs, while using encoders like LPG2vec to project graph topology into embeddings~\cite{pmlr-v198-besta22a}. The other treats NGDBs as latent-space data management systems, with neural storage and query engines operating over learned representations~\cite{ren2023neuralgraphreasoningcomplex}. Recent extensions further explore agentic~\cite{baitop} and privacy-preserving~\cite{hu2024privacy} NGDB architectures. However, existing evaluations provide limited coverage of realistic noise, expressive queries and continuous updates.


\textbf{Complex Logical Query Answering}. Complex query answering is important for NGDBs, especially under incomplete data. Prior work studies this problem through geometric query embeddings~\cite{HamiltonBZJL18,RenHL20,ZhangWCJW21}, logic-based reasoning~\cite{ArakelyanDMC21,ChenHS22}, inductive methods~\cite{Mikhail0001DWH22,Zhu0Z022}, and recent extensions to implicit constraints and abductive reasoning~\cite{yisen2025controllable,gao2025unifying}. Yet most remain limited to Existential First-Order logic, falling short of realistic database workloads with variable-length paths, aggregation, and data management operations.

\textbf{Existing Datasets and Limitations}. Existing benchmarks mainly target static knowledge-graph reasoning or predictive learning over relational databases. KG reasoning benchmarks, including LS-CQA~\cite{Bai0LYS24}, LitCQD~\cite{DemirWLNH23}, and $\text{EFO}{k}$-CQA~\cite{YinWFS23}, support attributes or richer logical queries but remain largely static and homogeneous. Relational benchmarks, such as RelBench v2~\cite{gu2026relbenchv2largescalebenchmark} and 4DBInfer~\cite{10.5555/3737916.3738772}, focus on structured prediction over heterogeneous relational data. They therefore fall short in evaluating expressive graph query execution, noise robustness, and dynamic data management.

%% file: sec/method.tex
\begin{figure*}[h]
    \centering
    \includegraphics[width=1\textwidth]{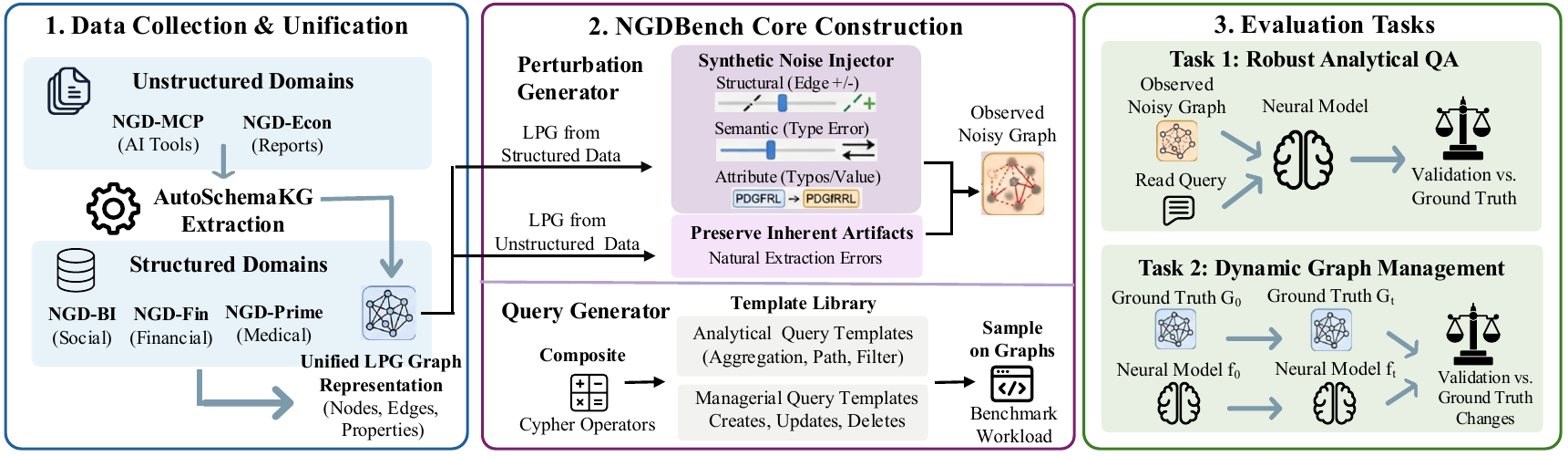}
    \vspace{-0.8cm}
    \caption{The NGDBench Framework Pipeline.(1) NGDBench unifies diverse data into LPG graphs, (2) constructs paired clean/noisy graphs via perturbation, and samples Cypher queries from template libraries. (3) Systems are evaluated on robust analytical QA and dynamic graph management under noisy observations.}
    \label{fig:framework}
    \vspace{-0.3cm}
\end{figure*}

\section{NGDBench}

We introduce \method{}, where each dataset therefore contains a clean graph representing latent ground truth, a corresponding perturbed graph representing observed data, and a collection of evaluation queries and management operations.

As illustrated in Fig. \ref{fig:framework}, our framework consists of three primary components. The \textbf{Data Collection \& Unification} module gathers representative structured and unstructured datasets, converting them into a unified graph representation. The \textbf{Perturbation Generator} module introduces controlled perturbations into the original graphs according to predefined specifications, thereby generating the observed graphs. Then, the \textbf{Query Generator} module constructs a \textit{Query Template Library} to sample both analytical and management-oriented queries. Lastly, comprehensive evaluation are implemented across robust analytical query to data management tasks.

\subsection{Data Preparation}

We construct \method{} from five domains: social systems, finance, biomedicine, agent tool use, and enterprise reporting. These domains cover both structured repositories and unstructured documents, reflecting realistic settings where data is heterogeneous, noisy, and only partially explicit. We prefix each dataset with \textit{NGD} to distinguish the processed benchmark graphs from their raw sources. Table~\ref{tab:summary} summarizes the dataset statistics.
\subsubsection{Structured Domains}

We first select three structured sources with diverse schemas, relation types, and query patterns. For each source, we convert the original data into a Labeled Property Graph while preserving its core semantics.

\noindent\textbf{NGD-BI.} Built on the LDBC-BI benchmark \cite{10.14778/3574245.3574270}, NGD-BI captures a complex social-network setting similar to those appearing in realistic web-agent environments \cite{zhou2024webarena}. It contains rich many-to-many relations, dense connectivity, and attribute-structure correlations, making it suitable for evaluating multi-hop reasoning and noise-robust relationship discovery.

\noindent\textbf{NGD-Fin.} Adapted from LDBC-FIN~\cite{10.14778/3746405.3746424}, NGD-Fin models financial transaction ecosystems involving accounts, entities, events, and monetary flows~\cite{yang2024finrobot}. Its temporal and numerical properties, together with multi-edge transaction patterns, make it suitable for evaluating analytical queries, latent-risk discovery, and robustness to noisy or missing financial relations.

\noindent\textbf{NGD-Prime.} Derived from PrimeKG \cite{chandak2022building}, NGD-Prime integrates biomedical entities such as diseases, drugs, genes, and biological processes. Its dense cross-type relations make it suitable for evaluating latent relation recovery and evidence aggregation across heterogeneous biomedical entities.

\subsubsection{Unstructured Domains}

We further include two unstructured sources to evaluate whether graph-based neural data management can expose latent structure from raw text. We use AutoSchemaKG \cite{bai2025autoschemakgautonomousknowledgegraph} to extract entities, relations, and schemas, and instantiate them as Labeled Property Graphs.

\noindent\textbf{NGD-MCP.} Built from MCP tool and trajectory data in Toucan-1.5M~\cite{xu2025toucan}, NGD-MCP captures agent workflow dependencies, supporting graph reasoning over tools, actions, and intermediate states.

\noindent\textbf{NGD-Econ.} NGD-Econ is derived from company annual reports in LongBench~\cite{bai2025longbenchv2deeperunderstanding}. Financial reports contain dispersed entities, metrics, events, and numerical evidence, which are important for long-document understanding and multi-step financial analysis \cite{chen2021finqa,reddy2024docfinqa}. By converting reports into graphs, NGD-Econ supports cross-document aggregation, relation discovery, and reasoning over implicit business and operational links.

\begin{table*}[t]
\centering
\scriptsize
\setlength{\tabcolsep}{2.5pt}
\renewcommand{\arraystretch}{0.95}

\caption{Detailed statistics of graph structures (groundtruth and perturbed graphs) and query categories across different domains.}
\vspace{-0.3cm}
\resizebox{\textwidth}{!}{
\begin{tabularx}{\textwidth}{p{1.9cm}p{2.5cm}YYcYYY}
\toprule

\multicolumn{2}{l}{\textbf{Category}} &
\textbf{NGD-BI} &
\textbf{NGD-Fin} &
\textbf{NGD-Prime} &
\textbf{NGD-MCP} &
\textbf{NGD-Econ} &
\textbf{Summary} \\

\midrule

\multirow{2}{*}{\shortstack[l]{Graph Structures\\(Groundtruth)}}
& \#Nodes
& 2,997,352
& 10,865
& 129,312
& 181,689
& 33,164
& -- \\

& \#Edges
& 17,196,776
& 57,818
& 8,100,498
& 351,263
& 63,757
& -- \\

\midrule

\multirow{2}{*}{\shortstack[l]{Graph Structures\\(Perturbed)}}
& \#Nodes
& 2,997,352
& 10,865
& 129,312
& 181,689
& 33,164
& -- \\

& \#Edges
& 17,971,365
& 60,476
& 8,465,124
& 351,263
& 63,757
& -- \\

\midrule

\multirow{3}{*}{Query Categories}
& \#Analytical (No Agg)
& 17,880
& 4,264
& 14,043
& 8,336
& 1,707
& 58,321 \\

& \#Analytical (Agg.)
& 9,580
& 4,412
& 6,494
& 4,167
& 1,050
& 25,703 \\


& \#Management
& 2,503
& 1,400
& 2,207
& --
& --
& 6,110 \\
\bottomrule
\end{tabularx}
}
\label{tab:summary}
\end{table*}

\subsection{Perturbation Generation}
\label{sec:perturbation_generation}

The perturbation component is designed to simulate noisy data conditions commonly found in real-world data management. To ground this design, we survey real-world datasets, prior synthetic-noise studies and data quality reports, with detailed evidence summarized in Appendix~\ref{sec:appendix-perturbation}, Table~\ref{tab:noise_survey_ranges}. The survey shows that noise levels vary substantially across domains and noise families: reported real-world noise ranges from 0.55\% attribute errors to over 70\% missing relations, while prior synthetic studies commonly inject noise at 10\%--50\%.

These observations suggest that no single fixed noise ratio can faithfully represent real-world data imperfections. We therefore design \method{} as a configurable and traceable perturbation environment rather than a fixed-noise benchmark. It covers the major noise families identified in the survey and exposes parameterized controls over the injection ratio of each noise type, enabling both moderate default evaluation and customized stress testing. For each generated observed graph, \method{} records the injected perturbations in noise log files, allowing users to inspect which nodes, edges, labels, or attributes were modified. Details on noise categories, real-world motivation, injection ratios, and perturbation logs are provided in Appendix~\ref{sec:appendix-perturbation}.

For the structured datasets (\textit{NGD-BI}, \textit{NGD-Fin}, and \textit{NGD-Prime}), we perturb clean latent graphs to construct observed graphs. The Perturbation Generator injects topological, schema, and attribute noise. These perturbations capture the gap between clean latent structure and imperfect observed data.

For the unstructured datasets (\textit{NGD-MCP} and \textit{NGD-Econ}), we do not add extra perturbations. Their graphs are produced by automated extraction pipelines and already contain realistic imperfections, such as entity duplication, mention variation, relation extraction errors, and type inference errors.


\subsection{Query Generation}
\label{sec:query_generation}

To align \textsc{NGDBench} with practical graph data management systems, we adopt the \textbf{Labeled Property Graph (LPG)} model and the \textbf{Cypher} query language. LPGs store rich properties directly on nodes and edges, making them suitable for real-world data with attributes on both entities and relations, such as financial transactions with timestamps and amounts. Cypher is the de facto query language for LPGs and is widely supported by modern graph engines such as Neo4j~\footnote{https://neo4j.com/} and FalkorDB~\footnote{https://www.falkordb.com/}, improving the benchmark's practical relevance.

We construct a Cypher query template library by organizing standard Cypher patterns into core operators, and generate benchmark queries using a perturbation-aware query sampler.
\subsubsection{Query Template Library Construction}

To ensure broad Cypher coverage, we survey both the official Cypher operator set\footnote{https://neo4j.com/docs/cypher-manual/25/planning-and-tuning/operators/operators-detail} and representative production-level Cypher queries. Although production queries can be complex, their logical structure is typically composed from recurring functional building blocks. Many official operators are runtime variants of these core functions, while others are not central to logical query semantics.

We therefore extract 29 core functional operators, including 23 analytical operators and 5 management operators, to guide query construction. These operators cover retrieval, traversal, filtering, projection, aggregation, and state-changing primitives. We then build analytical templates in stages, from atomic value-agnostic templates to no-aggregation (NoAGG) templates and aggregation (AGG) templates with \texttt{SUM}, \texttt{COUNT}, \texttt{MIN}, \texttt{MAX}, and \texttt{AVG}. This design captures the operator- and pattern-level foundations of many production analytical queries, making \method{} a starting point for future production-like neural data management workloads.

We further construct management templates to evaluate dynamic data management. These templates focus on frequent fine-grained operations, organized into batches that represent logical units of work involving creation, update, or deletion. Details of the operators, query templates, and management workloads are provided in Appendices~\ref{sec:appendix-operator}, \ref{sec:appendix-querytemplate}, \ref{sec:appendix-manage}, and \ref{sec:appendix-full_query_template}.
\vspace{-0.8em} 
\subsubsection{Query Sampling and Reconstruction}

We design a controllable query sampler that instantiates templates over both clean and perturbed graph regions. Users can specify the sampling ratio between clean regions, which support basic execution evaluation, and noisy regions, which test robustness under incomplete and corrupted observations.

We then reconstruct sampled Cypher queries into natural-language questions using an LLM, creating paired text-query instances for Text-to-Cypher and semantic understanding evaluation.

For non-aggregation queries with large answer sets, we reconstruct them as Boolean verification tasks rather than discarding them. Queries with fewer than 32 answers are kept unchanged; otherwise, we sample positive candidates from ground-truth answers and negatives from unrelated entities, and ask whether each candidate satisfies the query constraints. This setting better reflects practical scenarios where users verify specific entities rather than exhaustively enumerate all answers. Details are provided in Appendix~\ref{sec:boolean}.

\subsection{Task Formulation}
\label{sec:task_formulation}

We evaluate \textsc{NGDBench} along the following two dimensions.

\vspace{-0.8em}
\subsubsection{Task I: Robust Analytical Querying}

This task evaluates whether a model can answer analytical queries over imperfect observed data while recovering the intended latent information.

\paragraph{Problem Definition.}
Let $\mathcal{G}^*$ denote the clean latent graph, and let $\tilde{\mathcal{G}}=\Phi(\mathcal{G}^*)$ be the observed graph generated by a perturbation function $\Phi$. Given a natural language query $q$ and the observed graph context $\tilde{\mathcal{G}}$, a model $f_\theta$ produces either a direct answer or a formal query such as Cypher. The target answer is obtained by executing the query logic on the clean graph:
\begin{equation}
    a^* = \text{Exec}(q, \mathcal{G}^*).
\end{equation}
The objective is to minimize the discrepancy between the model prediction and the clean-graph answer:
\begin{equation}
    \min_\theta \mathcal{L}\left(f_\theta(q,\tilde{\mathcal{G}}), a^*\right),
\end{equation}
where $\mathcal{L}$ denotes the task-specific evaluation metric, such as F1 score, accuracy, or relative error. This task tests whether models can reason beyond noisy observations rather than merely querying the perturbed graph.

\subsubsection{Task II: Sequential Editing for Dynamic Data Management}

This task evaluates whether a model can track graph states under sequential \texttt{CREATE}, \texttt{UPDATE}, and \texttt{DELETE} operations.

\paragraph{Problem Definition.}
Let $\mathcal{G}_0$ be the initial graph state. At each step $t$ in a $T$-step session, the model receives a modification instruction $m_t$ and a validation query $q_t$. The graph update in Eq.~\eqref{eq:state-update}, the gold-answer computation in Eq.~\eqref{eq:gold-answer}, and the model prediction in Eq.~\eqref{eq:model-prediction} are defined as follows, where $\mathcal{H}_t=\langle \mathcal{G}_0,m_1,\ldots,m_t\rangle$ denotes the modification history:
\begin{align}
    \mathcal{G}_t &= \operatorname{Apply}(\mathcal{G}_{t-1}, m_t),
    \label{eq:state-update} \\
    a_t &= \operatorname{Exec}(q_t, \mathcal{G}_t),
    \label{eq:gold-answer} \\
    \hat{a}_t &= \operatorname*{argmax}_{a \in \mathcal{A}} 
    P_\theta(a \mid q_t,\mathcal{H}_t).
    \label{eq:model-prediction}
\end{align}

Then, we report step-wise accuracy over the full trajectory:
\begin{equation}
    \mathcal{L}_{step} = \frac{1}{T} \sum_{t=1}^{T} \mathbb{I}(\hat{a}_t = a_t),
\end{equation}
where $\mathbb{I}(\cdot)$ indicates whether the predicted answer matches the ground truth. This metric penalizes intermediate state-tracking errors rather than evaluating only the final state.

%% file: sec/experiment.tex
\begin{table*}[t]
\centering
\scriptsize  
\setlength{\tabcolsep}{2pt}  
\renewcommand{\arraystretch}{0.95} 
\setlength{\tabcolsep}{3pt}
\caption{Performance comparison across three datasets under non-aggregation and aggregation settings. (``——'' indicates values exceeding 10,000. Bold and \underline{underlined} values denote the best and second-best results respectively.)}
\vspace{-0.3cm}
\resizebox{\textwidth}{!}{
\begin{tabular}{l
cc ccc
cc ccc
cc ccc}
\toprule
\textbf{Baseline} &
\multicolumn{5}{c}{\textbf{NGD-Fin}} &
\multicolumn{5}{c}{\textbf{NGD-BI}} &
\multicolumn{5}{c}{\textbf{NGD-Prime}} \\
\cmidrule(lr){2-6} \cmidrule(lr){7-11} \cmidrule(lr){12-16}
&
\multicolumn{2}{c}{\textbf{NoAgg}} &
\multicolumn{3}{c}{\textbf{Agg}} &
\multicolumn{2}{c}{\textbf{NoAgg}} &
\multicolumn{3}{c}{\textbf{Agg}} &
\multicolumn{2}{c}{\textbf{NoAgg}} &
\multicolumn{3}{c}{\textbf{Agg}} \\
\cmidrule(lr){2-3} \cmidrule(lr){4-6}
\cmidrule(lr){7-8} \cmidrule(lr){9-11}
\cmidrule(lr){12-13} \cmidrule(lr){14-16}
&
\textbf{Jaccard} & \textbf{F1} &
\textbf{MdRE} & \textbf{MSLE} & \textbf{sMAPE} &
\textbf{Jaccard} & \textbf{F1} &
\textbf{MdRE} & \textbf{MSLE} & \textbf{sMAPE} &
\textbf{Jaccard} & \textbf{F1} &
\textbf{MdRE} & \textbf{MSLE} & \textbf{sMAPE} \\
\midrule
Oracle Cypher   & 0.448 & 0.471 & 0.039& 0.649 & 0.119 & 0.887 & 0.911 & 0.355 & 8.388 &  0.783& 0.736 & 0.790 & 0.203 &7.990  &0.562  \\
\cmidrule(lr){1-16}
Neo4j-Text2Cypher &  0.319& 0.319 & 0.660 & 31.877 & 0.975 & 0.212 &0.213  & —— &45.582  & 1.681 &  0.208 & 0.195 & —— & 40.760 & 1.481 \\
GPT5.1-Codex  &  \underline{0.358}& \underline{0.359} & 0.111 & 12.425 & 0.394 & \textbf{0.555} &\textbf{0.571}  & \textbf{0.987} & \textbf{15.921} & \textbf{1.089} &\textbf{0.510}  & \textbf{0.537} &\textbf{0.206} &\textbf{12.250}  & \textbf{0.743} \\
DeepseekV3.2  &  0.345& 0.345 & \textbf{0.102} &\textbf{5.636} & \underline{0.334} &\underline{0.394}  &\underline{0.404} & \underline{0.992} & \underline{20.862} & \underline{1.182} & \underline{0.426}& \underline{0.446} & \underline{0.409} & \underline{12.547} & \underline{0.835} \\
Qwen3-Coder &  \textbf{0.441}& \textbf{0.456} & \textbf{0.102} & \underline{6.054}& \textbf{0.332} & 0.306 & 0.315& \underline{0.992} & 22.655 & 1.198 & 0.389 & 0.410 & 0.430 & 13.197 & 0.852 \\
GraphRAG  & 0.004 & 0.004 & 161.500 & 27.770 & 1.564 & 0.001 & 0.001 & —— & 52.527 & 1.994 & 0.002 & 0.002 & 705.650 & 40.659 & 1.725 \\
\bottomrule
\end{tabular}
}
\label{tab:aggnoagg}
\end{table*}

\begin{table*}[t]
\centering

\scriptsize
\setlength{\tabcolsep}{6pt}
\caption{Performance comparison on management queries across different datasets.}
\vspace{-0.3cm}
\scalebox{0.85}{
\begin{tabular}{l
ccc
ccc
ccc}

\toprule
\textbf{Baseline} &
\multicolumn{3}{c}{\textbf{NGD-Fin}} &
\multicolumn{3}{c}{\textbf{NGD-BI}} &
\multicolumn{3}{c}{\textbf{NGD-Prime}} \\
\cmidrule(lr){2-4} \cmidrule(lr){5-7} \cmidrule(lr){8-10}
&
\textbf{MLRE} & \textbf{MSLE} & \textbf{sMAPE} &
\textbf{MLRE} & \textbf{MSLE} & \textbf{sMAPE} &
\textbf{MLRE} & \textbf{MSLE} & \textbf{sMAPE} \\
\midrule
Neo4j-Text2Cypher& 8.065 &17.038 &1.137  & 0.759 & 3.323 & 0.280 &3.667  &16.072  & \underline{0.337} \\
GPT5.1-Codex  &  3.036& \underline{11.727} & 0.771 & \underline{0.516} &\textbf{2.327}  & \underline{0.233} &\textbf{3.342}  &\textbf{15.138}  & 0.360 \\
DeepseekV3.2 & \underline{2.778} & 14.940 & \underline{0.716} & \textbf{0.455}& \underline{2.330} & \textbf{0.171} & 3.665  &  16.130 & 0.352  \\
Qwen3-Coder  & \textbf{2.676} & \textbf{4.619} & \textbf{0.610} & 0.562 & 2.656 &  0.239 & \underline{3.403}& \underline{15.619} & \textbf{0.309} \\
GraphRAG           & 5.129 & 17.510 & 1.518 & 9.043 & 81.106 & 1.979 & 6.337 & 41.776 & 1.958 \\
\bottomrule
\end{tabular}
}
\label{tab:manage}
\end{table*}

\section{Experiment}

\subsection{Analytical Query on Structured Domains}
\label{sec:sturcture}

\textbf{Experimental Settings and Metrics.}
For each benchmark query, we execute the oracle Cypher query on a traditional graph database Neo4j over both the clean latent graph and the perturbed observed graph. The clean-graph result serves as the target answer, while the observed-graph result provides an oracle upper bound for methods restricted to noisy observations.

We use translated natural-language questions as model inputs and evaluate three query settings. 
(1) \emph{No Aggregation (NoAgg)} queries return answer sets, so we report \textbf{Jaccard similarity} and \textbf{F1 score}.
(2) \emph{Boolean Queries} reformulate large-result queries as binary verification tasks, and we report \textbf{Accuracy}.
(3) \emph{Aggregation (Agg)} queries return scalar numerical values. To account for large variances in answer scales, we adopt relative-error-based metrics: \textbf{Median Relative Error (MdRE)} to capture typical prediction error while remaining robust to outliers, \textbf{Mean Squared Logarithmic Error (MSLE)} to penalize multiplicative deviations, and \textbf{Symmetric Mean Absolute Percentage Error (sMAPE)} for a scale-invariant and zero-robust measure of error. We have presented the calculation formula for metric in Appendix \ref{app:metrics}.


\textbf{Baselines.} We evaluate two representative neural-based graph query paradigms: \emph{Text-to-Cypher} and \emph{GraphRAG}. These are currently the most common ways LLM-based systems access and reason over graph data. Text-to-Cypher uses an LLM to convert query intent into executable graph queries, thereby combining neural language understanding with symbolic database execution. GraphRAG instead retrieves relevant graph evidence, typically in textualized triple form, and asks an LLM to reason over the retrieved context. Compared with exact symbolic matching, both paradigms have greater potential to tolerate noisy observations and capture latent semantic associations, making them suitable test cases for neural graph data management.

The \emph{Text-to-Cypher} baselines include the standard Neo4j-Text2Cypher implementation and strong code-generation LLMs, including DeepSeek V3.2~\cite{deepseekai2025deepseekv32pushingfrontieropen}, GPT-5.1-codex\footnote{https://openai.com/zh-Hans-CN/index/gpt-5-1-codex-max/}, and Qwen3-Coder-480B~\cite{yang2025qwen3technicalreport}. For \emph{GraphRAG}, we use a dense triple retriever with Qwen3-0.6B~\cite{qwen3technicalreport}, retrieving the top-15 triples by semantic similarity.

Traditional CQA methods are excluded because they are largely limited to Existential First-Order logic and cannot execute full Cypher workloads such as aggregation or dynamic operations. Generic vector databases are also excluded because they mainly support similarity search over pre-indexed embeddings. Handling noise distributed across labels, attributes, relations, and paths would require additional task-specific indexing and retrieval design, which falls outside their default query model. We instead include an \emph{Oracle Cypher} control on the perturbed graph to isolate data-noise effects from model generation errors.

\textbf{Results on NoAgg and Agg Queries.}
Table~\ref{tab:aggnoagg} reports results on NoAgg and Agg queries. Text-to-Cypher methods generally outperform GraphRAG, especially on aggregation queries, because structured query execution can retrieve complete matching results, while GraphRAG relies on dense top-$k$ retrieval and may miss relevant evidence when many triples are required.

However, Text-to-Cypher is still bounded by noise in the observed graph. Some aggregation-error metrics can become very large when realistic corruptions affect numerical attributes, since small data-entry or OCR may substantially distort downstream analytical results such as \texttt{SUM} or \texttt{AVG}. To separate model errors from data-noise effects, we split queries into consistent and inconsistent subsets, depending on whether their answers change after perturbation. Results in Appendix~\ref{app:moreexp} show that models perform much better on consistent queries but degrade sharply on inconsistent ones, indicating that the gap between observed data and latent ground truth is a key bottleneck.

Overall, Text-to-Cypher works when observations are reliable but fails to infer beyond corrupted or missing records. GraphRAG uses textual evidence flexibly but lacks complete recall and precise reasoning. These limitations motivate neural graph data management methods that combine symbolic execution with inference over noisy observations.

\begin{figure*}[h]
    \centering
    \includegraphics[width=0.9\textwidth]{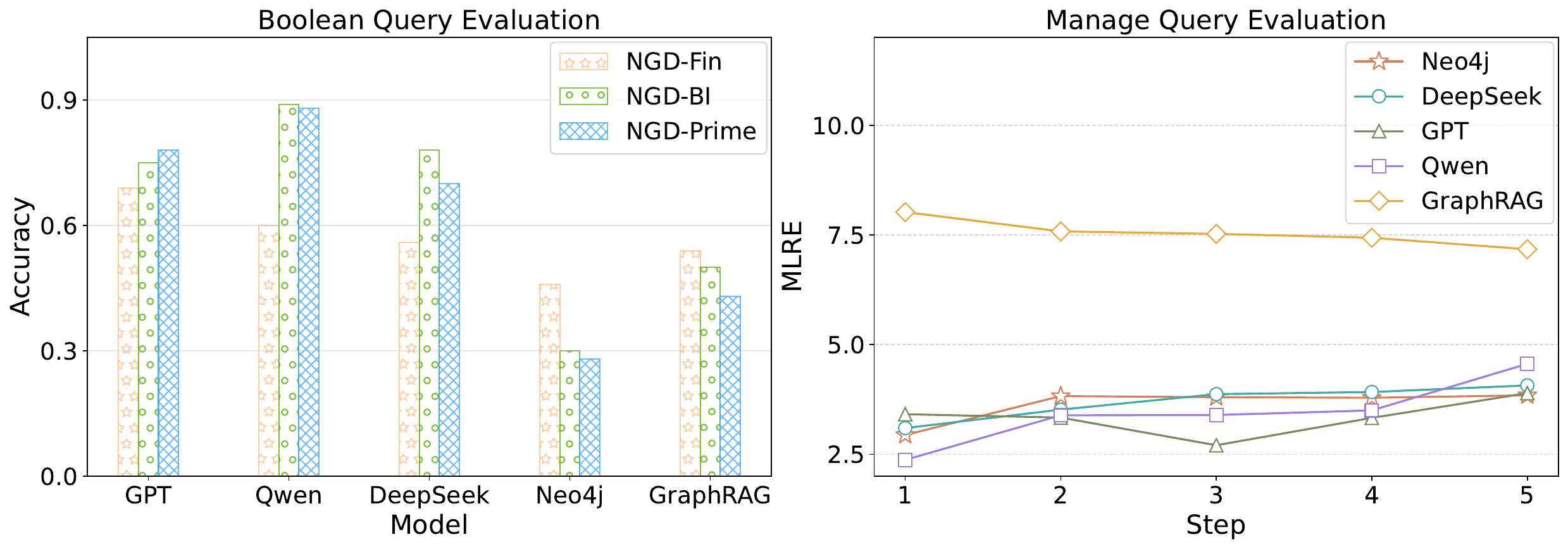}
    \vspace{-0.2cm}
    \caption{The left figure is the evaluation of boolean queries on three datasets. The right figure is the evaluation of dynamic steps on NGD-Prime, where lower MLRE indicates better performance.}
    \label{fig:step_evaluation}
\end{figure*}

\textbf{Results on Boolean Queries.}
Figure~\ref{fig:step_evaluation} left reports results on Boolean queries. In this setting, GraphRAG improves substantially and becomes competitive with Text-to-Cypher. The candidate-based formulation narrows the retrieval space, making dense retrieval more reliable. In contrast, Boolean verification can require more complex Cypher logic, making it less convenient for Text-to-Cypher models. This suggests that RAG-style methods can be effective for local verification, but remain insufficient for complex analytical queries requiring broad and complete evidence retrieval.

\subsection{Managerial Queries on Structured Data}

We further extend our evaluation to the dynamic management phase. In this setting, each management batch consists of five sequential steps; after each step, we execute a validation query to verify the updated state. Given the varying scales of the target values, we adopt a suite of relative-error-based metrics: \textbf{Mean Log Relative Error (MLRE)} is used to capture typical prediction errors while remaining robust to outliers; \textbf{Mean Squared Logarithmic Error (MSLE)} is employed to penalize multiplicative deviations rather than additive ones; and \textbf{Symmetric Mean Absolute Percentage Error (sMAPE)} is selected for its scale invariance and stability, particularly when handling values close to zero.

The results are shown in Table~\ref{tab:manage} and Fig.~\ref{fig:step_evaluation} right. Text-to-Cypher methods perform well because it can generate explicit edit operations that directly update the graph. However, it is prone to error accumulation: an incorrect update at an earlier step can propagate and affect subsequent states. In contrast, GraphRAG reasons over the full edit history in context, which helps mitigate cascading errors.

\subsection{Analytical Query on Unstructured Data}
\label{sec:unstructure}
We further evaluate graph-based RAG methods on the unstructured datasets, NGD-MCP and NGD-Econ. We compare three representative retrieval-based baselines: GraphRAG, which performs embedding-based retrieval over textualized triples extracted by our pipeline; HippoRAG2~\cite{gutierrez2025rag}, a recent text-first graph RAG method operating over raw documents; and ToG\cite{thinkongraph}, a hybrid retrieval framework that iteratively searches over graph and textual contexts. Following prior graph-based RAG studies~\cite{gutierrez2025rag,trivedi2022musique}, we report token-level F1 for non-aggregation queries. We report \textbf{MSLE} for aggregation queries, and more metrics on this task are desplayed in Appendix \ref{app:moreexp}, Table \ref{tab:rag-agg-metrics}.

As shown in Table~\ref{tab:unstructure}, GraphRAG achieves the best results across both datasets, obtaining the lowest aggregation error and the highest non-aggregation F1. This suggests that directly retrieving textualized triples provides stronger evidence than reconstructing query-specific reasoning graphs from raw passages, especially when questions are grounded in local graph neighborhoods. However, the absolute performance remains limited: even the best NoAgg F1 is below 0.25, and aggregation errors remain non-negligible, indicating that current RAG pipelines still struggle with precise complex querying and numerical reasoning.

This reflects a key limitation of retrieval-based pipelines: retrieval and computation are separated. RAG methods first retrieve a bounded top-$k$ context and then ask the LLM to reason over truncated evidence. While this can help when answers appear in a small local neighborhood, it is weak for complex path queries and aggregation tasks, where accurate answers require complete evidence rather than a few retrieved snippets. As a result, RAG-based methods may miss relevant tuples, include distractors, or fail to compute exact counts and aggregate values. We provide concrete success and failure cases in Appendix~\ref{case_study}.

\begin{table}[h]
\centering
\small
\setlength{\tabcolsep}{6pt}
\caption{Performance of graph-based RAG methods on unstructured datasets. Agg reports MSLE ($\downarrow$), and NoAgg reports F1 ($\uparrow$).}
\begin{tabular}{lcccc}
\toprule
\multirow{2}{*}{Baseline} & \multicolumn{2}{c}{NGD-MCP} & \multicolumn{2}{c}{NGD-Econ} \\
\cmidrule(lr){2-3} \cmidrule(lr){4-5}
 & Agg $\downarrow$ & NoAgg $\uparrow$ & Agg $\downarrow$ & NoAgg $\uparrow$ \\
\midrule
GraphRAG  & \textbf{0.169} & \textbf{0.194} & \textbf{0.673} & \textbf{0.242} \\
HippoRAG2 & 1.050 & 0.022 & 2.138 & 0.040 \\
ToG-RAG   & 0.359 & 0.013 & 3.809 & 0.212 \\
\bottomrule
\end{tabular}
\label{tab:unstructure}
\end{table}

%% file: sec/conclusion.tex
\section{Conclusion}

We introduced \method{}, a benchmark for advancing graph-based neural data management beyond passive querying over observed data. Unlike prior benchmarks limited to elementary logical operations, \method{} spans five diverse domains and supports the full spectrum of Cypher queries, while rigorously testing robustness against realistic noise and dynamic updates. Our evaluation of state-of-the-art LLMs and RAG approaches reveals significant limitations in precise analytical reasoning and noise resilience. Consequently, \method{} serves as a critical testbed for future research, paving the way for more reliable and capable neural database systems. We hope \method{} will stimulate principled work on noise-aware querying, controlling error accumulation in sequential edits, and standardized evaluation protocols for next-generation neural data management systems.

\newpage
\section*{Limitations}
One limitation of \method{} is that the dynamic management workload is currently constructed only for structured graph datasets. For graphs extracted from unstructured sources, such as documents or reports, we do not yet provide management queries involving \texttt{CREATE}, \texttt{UPDATE}, or \texttt{DELETE}. This is because editing extracted knowledge graphs is not simply a syntactic operation. It often requires factual grounding and domain-aware validation, especially in knowledge-intensive domains such as biomedicine, where adding, revising, or deleting a relation should be supported by reliable evidence.

In addition, our evaluation primarily focuses on answer accuracy under noisy and evolving graph states. Other practical factors, such as execution efficiency, memory cost, uncertainty calibration, and evidence quality, are left for future work. Extending \method{} along these dimensions would make it a more comprehensive testbed for neural graph data management.

Nevertheless, extending dynamic management workloads to unstructured-source graphs is an important direction. In real-world data management, unstructured data is also continuously evolving: new documents arrive, outdated information must be updated or archived, and extracted entities and relations may need to be corrected as evidence changes. Designing principled management operations for such graphs would enable \method{} to better evaluate how future neural data management systems maintain consistency over evolving, partially observed, and evidence-grounded knowledge.

%% file: sec/appendix.tex
\section{AI Use Statement}
We used AI tools solely to assist with writing refinement (e.g., polishing and editing). In addition, during dataset construction, we used a large language model to automatically generate natural-language versions of cypher queries. All instances where large language models were used in the methodology or dataset pipeline are explicitly disclosed in the main text.

\section{License and Data Usage Statement}
\label{sec:appendix-license}

The raw data sources used to construct \method{} are released under permissive licenses, including the MIT License and the Apache License 2.0. \method{} is a curated benchmark derived from these sources through schema normalization, graph construction, perturbation generation, and query annotation. These transformations are permitted under the corresponding licenses, and the resulting benchmark is released in compliance with their terms. We do not claim ownership of the original raw data sources, and users should follow the original licenses when redistributing, modifying, or extending \method{}.

\section{Details of Perturbation Generation}\label{sec:appendix-perturbation}

We design the perturbation module to approximate common data quality problems in the environments agents operate over. The goal is to evaluate whether a neural database can still answer queries and manage state when the observed graph differs from the latent graph because of missing records, extraction errors, schema ambiguity, or spurious links. Table~\ref{tab:noise_summary} summarizes the perturbation families used in \method{} and their default ratios. The default ratios are set according to Table \ref{tab:noise_survey_ranges}.

\begin{table*}[t]
\centering
\caption{Perturbation families used in \method{} and their default injection ratios.}
\label{tab:noise_summary}
\small
\setlength{\tabcolsep}{4pt}
\begin{tabularx}{\textwidth}{p{2.5cm} X X >{\centering\arraybackslash}p{1.8cm}}
\toprule
\textbf{Noise Family} & \textbf{What It Simulates} & \textbf{Typical Manifestation} & \textbf{Default Ratio} \\
\midrule
Incomplete edges & Missing facts or partial records & Removed relations in transactional, social, or biomedical graphs & 0.10 \\
False edges & Spurious joins or extraction mistakes & Incorrect links introduced by integration or inference pipelines & 0.15 \\
Relation-type noise & Ambiguous relation semantics & A valid edge assigned the wrong relation label & 0.20 \\
Node-type noise & Entity typing ambiguity & A node assigned an incorrect semantic type & 0.20 \\
Attribute noise & Local textual or numerical corruption & Typos, OCR artifacts, malformed values, or numeric deviations & 0.25 \\
\bottomrule
\end{tabularx}
\end{table*}

Our perturbation design is motivated by four broad categories of real-world noise. First, incompleteness and missing facts are common in both knowledge graphs and database integration pipelines \cite{yang2022rethinking,shi2018openworld,guo2023datacleaning}. Missing links may arise during extraction, digitization, or data fusion, and are widely recognized as a core challenge under the open-world assumption \cite{yang2022rethinking,shi2018openworld}. This is captured in \method{} by \textit{incomplete\_edges}. In NGD-Fin, for example, we remove selected edges between companies and loans to simulate omitted financial records or missing loan agreements in corporate filings and transaction logs \cite{bryzgalova2025missing}.

Second, erroneous attribute values are pervasive in practice, arising from human data-entry mistakes, OCR failures, sensor malfunctions, transmission errors, and normalization inconsistencies \cite{guo2023datacleaning,althubaiti2016infobias}. These issues are reflected in our \textit{attribute\_noise} family. In NGD-Prime, for instance, we perturb the drug name \textit{Ornipressin} to \textit{Ornjpressin}, which mimics common spelling errors in medical records and drug information systems \cite{senger2010misspellings}.

Third, schema and type inconsistency frequently appears in noisy graph construction, especially when data are crowdsourced, automatically extracted, or integrated across heterogeneous ontologies \cite{jarnac2025uncertainty,zhao2026wikidata}. Typical manifestations include incorrect type labels, relation-type confusion, and semantic inconsistency between concept hierarchies. These issues are captured through \textit{relation\_type\_noise} and \textit{node\_type\_noise}. In NGD-Prime, for example, the entity \textit{Recurrent fractures} can be relabeled from \textit{effect/phenotype} to \textit{exposure}, simulating domain-specific misclassification caused by conceptual ambiguity in collaborative biomedical knowledge sources \cite{zhao2026wikidata,althubaiti2016infobias}.

Fourth, erroneous links are common in information extraction and data fusion pipelines, where systems may incorrectly join entities or create spurious relationships from ambiguous context \cite{jarnac2025uncertainty}. This corresponds to our \textit{false\_edges} perturbation family. In NGD-BI, for example, we insert spurious \textit{Comment\_isLocatedIn\_Country} edges between comments and unrelated countries to mimic realistic social-media analytics errors such as incorrect geolocation or ambiguous entity matching.

During the design of this module, we also surveyed reported noise characteristics in real-world datasets and synthetic noise settings used in prior graph denoising and refinement work \cite{yang2022rethinking,shi2018openworld,guo2023datacleaning,qian2023rtgnn,bryzgalova2025missing}. Table~\ref{tab:noise_survey_ranges} summarizes the main ranges we observed.

\begin{table*}[t]
\centering
\caption{Survey summary of noise characteristics reported in real-world datasets and prior synthetic-noise studies, compiled from \cite{yang2022rethinking,shi2018openworld,guo2023datacleaning,qian2023rtgnn,bryzgalova2025missing}. The ranges show that realistic noise levels vary substantially across domains and noise families.}
\label{tab:noise_survey_ranges}
\small
\begin{tabularx}{\textwidth}{l YYYY}
\toprule
\textbf{Source} & \textbf{Erroneous Attribute} & \textbf{Incompleteness} & \textbf{Type Inconsistency} & \textbf{Erroneous Links} \\
\midrule
Real-world datasets & 0.55\%--26.9\% in reported medical settings & Over 70\% missing \textit{born\_in}; about 90\% missing \textit{ethnicity} in Freebase-style settings & 8.0\%--38.5\% across reported semantic domains & 10\%--25\% in reported financial settings \\
Prior synthetic injection & 10\%--50\% & 2.9\% and 10.4\% in representative studies & 10\%--50\% & 10\%, 20\%, and 30\% \\
\bottomrule
\end{tabularx}
\end{table*}

Two observations motivate our design. First, real-world noise is highly variable and often skewed across noise families, so a single fixed ratio cannot capture all realistic settings. Second, prior denoising work typically evaluates only a small set of synthetic noise levels. We therefore expose perturbation as a configurable environment generator rather than commit to one universal corruption pattern.

Our default configuration sets \textit{incomplete\_edges} to 0.10, \textit{false\_edges} to 0.15, \textit{relation\_type\_noise} and \textit{node\_type\_noise} to 0.20, and \textit{attribute\_noise} to 0.25. These values fall within the ranges summarized in Table~\ref{tab:noise_survey_ranges} and define a moderate baseline setting that is neither trivial nor excessively destructive.

\section{Details of Query Generation Module}\label{sec:appendix-a}
\subsection{Query Operators Supported in \method{} }\label{sec:appendix-operator}
The operators used in \method{} is a minimized set covering all Cypher functionalities while ensuring runtime compatibility. For instance, the \textit{HashJoin} operator acts as a unified abstraction for variants like \textit{Node Hash Join}, \textit{Value Hash Join}, and \textit{Node Left/Right Outer Hash Join}. Operators strictly for performance optimization, such as \textit{Cache Properties}, were excluded as they do not influence query logic.
The operators are shown in Table \ref{tab:core_operators}.
\begin{table*}[h]\label{t6}
\centering
\footnotesize 
\caption{Classification and descriptions of the supported Cypher core operators serving as the benchmark's functional primitives.}
\vspace{-0.3cm}
\label{tab:core_operators}

\renewcommand{\arraystretch}{0.9} 
\setlength{\tabcolsep}{4pt}       

\begin{tabularx}{\textwidth}{l >{\raggedright\arraybackslash}p{4.3cm} X} 
\toprule
\textbf{Category} & \textbf{Core Operators} & \textbf{Function Description} \\
\midrule

\textbf{Data Access} 
 & \texttt{Scan}, \texttt{Seek} 
 & Entry operators (e.g., node scan, index seek). \\
\midrule

\textbf{Traversal \& Matching} 
 & \texttt{Expand}, \texttt{VarLengthExpand}, \texttt{ShortestPath}, \texttt{Repeat}, \texttt{Optional}, \texttt{Anti}, \texttt{Triadic} 
 & Topology navigation, variable-length paths, and pattern matching (incl. anti/optional). \\
\midrule

\textbf{Join \& Set Operations} 
 & \texttt{HashJoin}, \texttt{Apply}, \texttt{CartesianProduct}, \texttt{Argument}, \texttt{Union} 
 & Combine intermediate results (hash / nested loop); includes unions. \\
\midrule

\textbf{Filter \& Projection} 
 & \texttt{Filter}, \texttt{Project}, \texttt{Distinct} 
 & Predicates, expressions, field selection, and deduplication. \\
\midrule

\textbf{Aggregation \& Control} 
 & \texttt{Aggregate}, \texttt{Unwind}, \texttt{Sort}, \texttt{Limit}, \texttt{Skip}, \texttt{Top} 
 & Grouping, list flattening, and output control (order/limit). \\
\midrule

\textbf{Data Modification} 
 & \texttt{Create}, \texttt{Merge}, \texttt{Delete}, \texttt{Set}, \texttt{Foreach} 
 & Graph mutations: creation, deletion, property updates. \\

\bottomrule
\end{tabularx}
\end{table*}

\subsection{Analytical Query Generation}\label{sec:appendix-analyse}
The generation of our analytical query templates is structured into three progressive phases. The detailed workflow and corresponding example query templates for each phase are presented in Sections B2.2, B2.3, and B2.4, respectively.
\subsubsection{Query Template Design Principle}\label{sec:appendix-querytemplate}
When designing our query generation module, we surveyed Cypher and practical graph-database workloads such as LDBC FinBench. We found that, although production queries are diverse, they are typically compositional: complex queries can be decomposed into a relatively small set of core operators and recurring query patterns. Neo4j's documentation likewise presents Cypher as compositions of clauses and execution steps.

We therefore extracted 29 core functional operators and constructed templates that cover the main basic and complex query families, including chain, variable-length traversal, join, pipeline, anti-pattern, union, triadic, and aggregation patterns, rather than attempting to enumerate every production query verbatim. Our template library directly reflects this coverage.

For example, consider the following production-style query:

\begin{lstlisting}[basicstyle=\ttfamily\small,breaklines=true,breakatwhitespace=false]
MATCH (person:Person {id:8796093022538})-[edge1:own]->(pAcc:Account)
      -[edge2:transfer]->(compAcc:Account)<-[edge3:own]-(com:Company)
WHERE edge2.timestamp > 1635724800000 AND edge2.timestamp < 1638144000000
WITH compAcc.id AS compAccountId, sum(edge2.amount) AS sumEdge2Amount
RETURN compAccountId, round(sumEdge2Amount * 1000) / 1000 as sumEdge2Amount
ORDER BY sumEdge2Amount DESC;
\end{lstlisting}

Although this query appears complex as a whole, it can still be decomposed into the same building blocks already covered in NGDBench, such as \emph{edge-property filtering}, a pipeline stage using \texttt{WITH}, \emph{aggregation} via \texttt{sum}, \emph{projection/post-processing} in \texttt{RETURN}, and \emph{ordering} over the aggregated result.

As a result, queries seen in production can often be decomposed into the operator- and pattern-level building blocks already covered by NGDBench.
\subsubsection{Phase 1: Basic Template Synthesis}
\textbf{Description:} 
This phase utilizes core operators (\texttt{Scan}, \texttt{Seek}, \texttt{Expand}, \texttt{Projection}.) to construct the most fundamental query skeletons. These templates contain placeholders (e.g., \texttt{\$LABEL}, \texttt{\$PROP}) intended for subsequent parameter instantiation, corresponding to the \textit{Abstract Prototypes} in the algorithm.

\begin{lstlisting}[
  caption={Phase 1 Examples: Basic Lookup and Traversal},
  breaklines=true,
  breakatwhitespace=true,
  columns=fullflexible,
  basicstyle=\ttfamily\small
]
// Example 1: Basic Node Lookup
// Operators: Scan + Filter + Project
MATCH (n:$LABEL)
WHERE n.$PROP = $VALUE
RETURN n

// Example 2: 1-Hop Traversal
// Operators: Expand + Match
MATCH (n:$LABEL)-[:$REL_TYPE]->(m)
RETURN m
\end{lstlisting}

\subsubsection{Phase 2: Complexity Augmentation (NoAgg Set)}
\textbf{Description:} 
In this phase, basic templates are extended to include more complex logic without aggregation. This includes Path Logic (\texttt{VarLengthExpand}, \texttt{Triadic}), Relational Logic (\texttt{UNION}, \texttt{Cartesian Product}, \texttt{HashJoin}, \texttt{Repeat}), and complex filtering (\texttt{Distinct}, \texttt{Unwind}, \texttt{Apply}). The generated templates belong to the $\mathcal{T}_{NoAgg}$ set.

\begin{lstlisting}[
  caption={Phase 2 Examples: Path Logic and Set Operations},
  breaklines=true,
  breakatwhitespace=true,
  basicstyle=\ttfamily\small
]
// Example 1: Variable-Length Path with Filtering
// Operators: VarLengthExpand + Filter
MATCH p = (n:$LABEL)-[:$REL_TYPE*1..3]->(m)
WHERE n.age > $AGE_LIMIT AND m.status = 'Active'
RETURN p

// Example 2: Union Query
// Operators: Union
MATCH (n:$LABEL_A) WHERE n.type = 'Gold' RETURN n
UNION
MATCH (n:$LABEL_B) WHERE n.credit > $CREDIT RETURN n
\end{lstlisting}

\subsubsection{Phase 3: Aggregation Extension (Agg Set)}
\textbf{Description:} 
This phase builds upon the complex templates from Phase 2 by adding aggregation operators (such as \texttt{Sum}, \texttt{Min}, \texttt{Max}, \texttt{Avg}, \texttt{Count}). This allows for testing the system's performance when handling grouping and statistical calculations.

\begin{lstlisting}[
  caption={Phase 3 Examples: Statistical Aggregation},
  breaklines=true,
  breakatwhitespace=true,
  columns=fullflexible,
  basicstyle=\ttfamily\small
]
// Example 1: Neighbor Count Statistics
// Operators: Count + Group By (Implicit)
MATCH (n:$LABEL)-[:$REL_TYPE]->(m)
RETURN n.id, count(m) AS degree

// Example 2: Average Property Calculation
// Operators: Avg + Filter
MATCH (n:$LABEL)
WHERE n.created_year > $YEAR
RETURN avg(n.salary) AS avg_salary
\end{lstlisting}

\subsubsection{Boolean Query}\label{sec:boolean}
We filter the generated templates to identify those utilizing the \textit{Collect} or \textit{Cartesian Product} operator, which typically yield multi-valued list results. For each identified template (Positive Template), we synthesize a corresponding Inverse Query Template (e.g., using negation or anti-pattern logic). These are coupled to form a Positive-Inverse pair, creating a dedicated set for validating query engine consistency.
For instance, considering a scenario involving a Cartesian Product between Person nodes based on their IP locations, the generated pair is illustrated as follows:

\begin{itemize}

\item \textbf{Positive Template ($t_{pos}$):} 
Retrieves pairs of distinct persons sharing the \textit{same} location IP.

\begin{lstlisting}[
  caption={Template Examples: Positive vs Inverse Conditions},
  breaklines=true,
  breakatwhitespace=true,
  columns=fullflexible,
  basicstyle=\ttfamily\small
]
// Positive Template (t_pos)
// Retrieves pairs of distinct persons sharing the same location IP
MATCH (a:Person), (b:Person)
WHERE a.locationIP = b.locationIP AND a <> b
WITH a, b
RETURN a, collect(b.birthday)[0..1] AS bs

// Inverse Template (t_inv)
// Retrieves pairs of distinct persons with different location IPs
MATCH (a:Person), (b:Person)
WHERE a.locationIP <> b.locationIP AND a <> b
WITH a, b
RETURN a, collect(b.birthday)[0..1] AS bs
\end{lstlisting} 
.

\begin{lstlisting}[style=cypherstyle]
MATCH (a:Person), (b:Person)
WHERE a.locationIP <> b.locationIP AND a <> b
WITH a, b
RETURN a, collect(b.birthday)[0..1] AS bs
\end{lstlisting}

\end{itemize}

\subsection{Management Queries}\label{sec:appendix-manage}
We developed the \textbf{Management Query Template Library} to evaluate the system's transactional capabilities and data integrity under mutation. The generation workflow proceeds from Template Construction to Query Sampling:
\begin{enumerate}
\item \textbf{Template Construction}: We first define parameterized query skeletons (e.g., \texttt{CREATE (n:Person {id: \$id})}) covering standard graph mutation logic.
\item \textbf{Parameter Instantiation}: To generate executable queries, we sample concrete values for these parameters. Data generators inject valid IDs and attributes to ensure the queries are semantically valid against the current graph schema.
\item \textbf{Batch Assembly}: Finally, these instantiated query instances are organized into Batches. Each batch represents a logical unit of work designed to simulate real-world transactional sequences.
\end{enumerate}
A single batch is strictly structured as a sequence of $K$ steps (standard configuration $K=5$), with each step constituting an Operation-Validation Pair.
This pair consists of an Operation Template, which executes a mutation using operators including Create , Delete, Set, and Remove, to modify the graph state, and a corresponding Validation Template, a read-only query designed to immediately verify the effect of the preceding operation (e.g., confirming a node deletion). 

\textbf{Operation Template ($t{op}$)}: A mutation query template utilizing operators such as Create, Delete, Set, or Remove. This template modifies the graph state.

\textbf{Validation Template ($t{val}$)}: A corresponding read-only query template designed to verify the immediate effect of the preceding operation.

\textbf{Uniform Operation Batches}
These batches consist of repeated executions of the same operator type to test specific write throughput and stability.
\vspace{0.2em}
\begin{lstlisting}[
  caption={Uniform Batch Example},
  breaklines=true,
  breakatwhitespace=true,
  columns=fullflexible,
  basicstyle=\ttfamily\small
]
// Management Queries
CREATE (n:Person {city: 'Kaohsiung'})
      -[:Guarantee]->
      (g:Person {id: 21..})

CREATE (n:Person {city: 'Hanover'})
      -[:Guarantee]->
      (g:Person {id: 50})

CREATE (n:Person {city: 'Hamamatsu'})
      -[:Guarantee]->
      (g:Person {id: 21..})

CREATE (n:Person {city: 'Shanghai'})
      -[:Guarantee]->
      (g:Person {id: 911})

CREATE (n:Person {city: 'Manipal'})
      -[:Guarantee]->
      (g:Person {id: 21..})

// Validation Query
MATCH (g:Person)<-[:Guarantee]-(n:Person)
RETURN count(n)
\end{lstlisting}
\textit{Explanation:} The management queries sequentially insert five new Guarantee relationships between Person nodes from different cities. The validation query counts the total number of such relationships to verify that all five insertions were successfully persisted.

\vspace{0.8em}
\textbf{Mixed Operation Batches}
These batches interleave different types of operators (e.g., insertion followed by deletion) to evaluate the system's handling of complex transactional logic and isolation.
\vspace{0.2em}
\begin{lstlisting}[
  caption={Mixed Batch Example},
  breaklines=true,
  breakatwhitespace=true,
  columns=fullflexible,
  basicstyle=\ttfamily\small
]
// Management Queries
CREATE (n:TagClass {name: 'SportsMember'})
CREATE (n:TagClass {name: 'Company'})
MATCH (n:TagClass {id: 'SportsMember'})
DETACH DELETE n
CREATE (n:TagClass {name: 'Country'})
CREATE (n:TagClass {name: 'MemberOfParliament'})
MATCH (n:TagClass {id: 'Country'})
DETACH DELETE n

// Validation Query
MATCH (n:Post)
RETURN count(n) AS cnt
\end{lstlisting}
\textit{Explanation:} This sequence performs a "create-then-delete" stress test. It creates four new TagClass nodes but immediately deletes two of them ('SportsMember' and 'Country') within the same batch. The validation query checks the count of Post nodes, serving as a background integrity check to ensure these schema modifications do not corrupt unrelated data entities.

\section{Full Query Templates}\label{sec:appendix-full_query_template}
\subsection{No-aggregation Template}
This section lists the full non-aggregation analytical query templates used to instantiate the $\mathcal{T}_{\text{NoAgg}}$ library. We group templates by operator pattern to make the construction process easier to inspect.

\subsubsection{Basic Templates}
\begin{lstlisting}[style=cypherstyle]
T001:
MATCH (n:$LABEL {$PROP: $VALUE})-[:$REL_TYPE]->(m)
RETURN m

T002:
MATCH (a:$LABEL1)-[:$REL1]->(b:$LABEL2)
MATCH (a)-[:$REL2]->(b)
RETURN a, b

T003:
MATCH (a:$LABEL1)-[:$REL]-(b:$LABEL2)
WHERE b.$PROP $OP $VALUE
RETURN a
\end{lstlisting}
\subsubsection{Anti Templates}
\begin{lstlisting}[style=cypherstyle]
T001:
MATCH (a:$LABEL1 {$PROP1: $VALUE})-[:$REL_TYPE]-()-[:$REL_TYPE]-(b:$LABEL2)
WHERE NOT (a)-[:$REL_TYPE]-(b)
RETURN b.$RETURN_PROP

T002:
MATCH (a:$LABEL)-[:$REL2]->(:$LABEL3)
WHERE NOT (a)-[:$REL1]->(:$LABEL2)
RETURN a.$RET_PROP
\end{lstlisting}

\subsubsection{Union Templates}
\begin{lstlisting}[style=cypherstyle, caption={Union Templates}]
T001:
MATCH (a:$LABEL1 {$PROP1: $VALUE})
RETURN a.$PROP1
UNION
MATCH (b:$LABEL2 {$PROP2: $VALUE2})
RETURN b.$PROP2
\end{lstlisting}

\subsubsection{Chain Templates}
\begin{lstlisting}[style=cypherstyle, caption={Chain and Path Templates}]
T001:
MATCH (a:$LABEL1)-[:$REL1]->(b:$LABEL2)-[:$REL2]->(c:$LABEL3)-[:$REL3]->(d:$LABEL4)
RETURN a, b, d

T002:
MATCH (a:$LABEL1)-[:$REL_TYPE*$MIN_HOPS..$MAX_HOPS]-(b:$LABEL2)
RETURN a, b

T003:
MATCH (start:$START_LABEL {$START_PROP: $START_VALUE})
MATCH p = (start)
  ((n1:$L1)<-$D1[:$R1]-$D2(r:$REL_NODE_LABEL)-$D3[:$R2]->(n2:$L2)
  WHERE r.$NODE_PROP = $NODE_VALUE){$MIN_HOPS..$MAX_HOPS}
  (end:$END_LABEL)
RETURN p
\end{lstlisting}

\subsubsection{Pipelined Templates}

\begin{lstlisting}[style=cypherstyle, caption={Pipelined Templates with \texttt{WITH}}]
T001:
MATCH (a:$LABEL1 {$PROP1: $VALUE})-[:$REL1]->(b:$LABEL2)
WITH b
MATCH (b)-[:$REL2]->(c:$LABEL3)
RETURN b.$RET_PROP1, c.$RET_PROP2

T002:
MATCH (a:$LABEL1 {$PROP1: $VALUE})-[:$REL1]->(b:$LABEL2)
WITH b
MATCH (b)-[:$REL2]->(c:$LABEL3)
WITH b, c
MATCH (c)-[:$REL3]->(d:$LABEL4)
RETURN b.$RET_PROP1, c.$RET_PROP2, d.$RET_PROP3

T003:
MATCH (a:$LABEL1 {$PROP1: $VALUE})-[:$REL1]->(b:$LABEL2)
WITH b
MATCH (b)-[:$REL2]->(c:$LABEL3)
WHERE c.$FILTER_PROP = $FILTER_VAL
WITH b, c
MATCH (c)-[:$REL3]->(d:$LABEL4)
RETURN b.$RET_PROP1, c.$RET_PROP2, d.$RET_PROP3
\end{lstlisting}

\subsubsection{Nested-Loop Templates}
\begin{lstlisting}[style=cypherstyle, caption={Nested-Loop and Predicate Templates}]
T001:
MATCH (a:$LABEL1 {$PROP1: $VALUE})
MATCH (b:$LABEL2 {$PROP2: a.$REF_PROP})
WHERE a <> b
RETURN a, b

T002:
MATCH (a:$LABEL1)
WHERE (a)-[:$REL_TYPE]->(:$LABEL2)
RETURN a.$RETURN_PROP

T004:
MATCH (n:$LABEL)
WHERE (n)-[:$REL1]->(:$L1) OR (n)-[:$REL2]->(:$L2)
RETURN n.$PROP

T005:
MATCH (n:$LABEL)
WHERE NOT (n)-[:$REL1]->(:$L1) OR (n)-[:$REL2]->(:$L2)
RETURN n.$PROP

T006:
MATCH (n:$LABEL)
WHERE n.$PROP $OP $VAL OR (n)-[:$REL]->(:$L2)
RETURN n.$RET
\end{lstlisting}

\subsection{Aggregation Template}
\subsubsection{Basic}
\begin{lstlisting}[style=cypherstyle]
T001:
MATCH (n:$LABEL) RETURN sum(n)
\end{lstlisting}

\subsubsection{Basic Aggregation}
\begin{lstlisting}[style=cypherstyle]
T001:
MATCH (a:$LABEL) RETURN avg(a.$PROP) AS avg_value

T003:
MATCH (a:$LABEL) RETURN sum(a.$PROP) AS total_value

T004:
MATCH (a:$LABEL) RETURN min(a.$PROP) AS min_value

T005:
MATCH (a:$LABEL) RETURN max(a.$PROP) AS max_value
\end{lstlisting}

\subsubsection{Basic Expand Aggregation}
\begin{lstlisting}[style=cypherstyle]
T001:
MATCH (n:$LABEL {$PROP: $VALUE})-[:$REL_TYPE]->(m) RETURN count(m) AS cnt

T002:
MATCH (n:$LABEL {$PROP: $VALUE})-[:$REL_TYPE]->(m:$LABEL2) RETURN avg(m.$RET_PROP2) AS avg_value

T003:
MATCH (n:$LABEL {$PROP: $VALUE})-[:$REL_TYPE]->(m:$LABEL2) RETURN sum(m.$RET_PROP2) AS total
\end{lstlisting}

\subsubsection{Basic Expand Into Aggregation}
\begin{lstlisting}[style=cypherstyle]
T001:
MATCH (a:$LABEL1)-[:$REL1]->(b:$LABEL2) MATCH (a)-[:$REL2]->(b) RETURN count(*) AS cnt

T002:
MATCH (a:$LABEL1)-[:$REL1]->(b:$LABEL2) MATCH (a)-[:$REL2]->(b) RETURN a, count(b) AS cnt
\end{lstlisting}

\subsubsection{Basic Filter Aggregation}
\begin{lstlisting}[style=cypherstyle]
T001:
MATCH (a:$LABEL1)-[r:$REL_TYPE]-(b:$LABEL2) WHERE r.$REL_PROP $OP $VALUE RETURN count(a) AS cnt

T002:
MATCH (a:$LABEL1)-[r:$REL_TYPE]-(b:$LABEL2) WHERE r.$REL_PROP $OP $VALUE RETURN a, count(b) AS cnt

T003:
MATCH (a:$LABEL1)-[r:$REL_TYPE]-(b:$LABEL2) WHERE r.$REL_PROP $OP $VALUE RETURN avg(r.$REL_PROP), max(r.$REL_PROP)
\end{lstlisting}

\subsubsection{Nested Loop}
\begin{lstlisting}[style=cypherstyle]
T007:
MATCH (a:$LABEL1 {$PROP1: $VALUE}) MATCH (b:$LABEL2 {$PROP2: a.$REF_PROP}) RETURN count(b) AS cnt

T009:
MATCH (a:$LABEL1) WHERE (a)-[:$REL_TYPE]->(:$LABEL2) RETURN count(a) AS cnt

T011:
MATCH (a:$LABEL1 {$PROP1: $VALUE}), (b:$LABEL2) WHERE NOT (a)-[:$REL_TYPE]->(b) RETURN count(b) AS cnt

T012:
MATCH (n:$LABEL) WHERE (n)-[:$REL1]->(:$LABEL2) OR (n)-[:$REL2]->(:$LABEL3) RETURN count(n) AS cnt

T013:
MATCH (n:$LABEL) WHERE NOT (n)-[:$REL1]->(:$LABEL2) OR (n)-[:$REL2]->(:$LABEL3) RETURN count(n)

T014:
MATCH (n:$LABEL) WHERE n.$PROP $OP $VAL OR (n)-[:$REL]->(:$LABEL2) RETURN avg(n.$RET)
\end{lstlisting}

\subsubsection{Chain}
\begin{lstlisting}[style=cypherstyle]
T004:
MATCH (a:$LABEL1)-[:$REL1]->(b:$LABEL2)-[:$REL2]->(c:$LABEL3)-[:$REL3]->(d:$LABEL4) RETURN count(d) AS cnt

T005:
MATCH (a:$LABEL1)-[:$REL1]->(b:$LABEL2)-[:$REL2]->(c:$LABEL3)-[:$REL3]->(d:$LABEL4) RETURN a, count(d) AS cnt

T006:
MATCH (a:$LABEL1)-[:$REL1]->(b:$LABEL2)-[:$REL2]->(c:$LABEL3)-[:$REL3]->(d:$LABEL4) RETURN avg(d.$RET)

T008:
MATCH (a:$LABEL1)-[:$REL_TYPE *$MIN_HOPS..$MAX_HOPS]-(b:$LABEL2) RETURN count(b) AS cnt

T009:
MATCH (a:$LABEL1)-[:$REL_TYPE *$MIN_HOPS..$MAX_HOPS]-(b:$LABEL2) RETURN a, count(b) AS cnt

T010:
MATCH p = (a:$LABEL1)-[:$REL_TYPE *$MIN_HOPS..$MAX_HOPS]-(b:$LABEL2) RETURN avg(length(p)) AS avg_len

T011:
MATCH (start:$START_LABEL {$START_PROP: $START_VALUE}) MATCH p = (start) ((n1:$LABEL2)<-$D1[:$R1]-$D2(r:$REL_NODE_LABEL)-$D3[:$R2]->(n2:$LABEL3) WHERE r.$NODE_PROP = $NODE_VALUE){$MIN_HOPS..$MAX_HOPS} (end:$END_LABEL) RETURN count(p) AS path_count

T012:
MATCH (start:$START_LABEL {$START_PROP: $START_VALUE}) MATCH p = (start) (...){$MIN_HOPS..$MAX_HOPS} (end:$END_LABEL) RETURN avg(length(p)) AS avg_len
\end{lstlisting}

\subsubsection{Join}
\begin{lstlisting}[style=cypherstyle]
T004:
MATCH (a:$LABEL1), (b:$LABEL2) WHERE a.$PROP1 = b.$PROP2 AND a <> b RETURN count(*) AS cnt

T006:
MATCH (a:$LABEL1), (b:$LABEL2) WHERE a.$PROP1 = b.$PROP2 AND a <> b WITH a, count(b) AS cnt WHERE cnt > $THRESHOLD RETURN a
\end{lstlisting}

\subsubsection{Pipelined}
\begin{lstlisting}[style=cypherstyle]
T001:
MATCH (a:$LABEL1 {$PROP1: $VALUE})-[:$REL1]->(b:$LABEL2) WITH b, count(*) AS cnt WHERE cnt > 0 RETURN b.$RET_PROP1

T002:
MATCH (a:$LABEL1 {$PROP1: $VALUE})-[:$REL1]->(b:$LABEL2) WITH b MATCH (b)-[:$REL2]->(c:$LABEL3) WHERE c.$FILTER_PROP = $FILTER_VAL RETURN count(*) AS total_cnt

T003:
MATCH (a:$LABEL1 {$PROP1: $VALUE})-[:$REL1]->(b:$LABEL2) WITH b MATCH (b)-[:$REL2]->(c:$LABEL3) WHERE c.$FILTER_PROP = $FILTER_VAL RETURN b, count(c) AS cnt

T004:
MATCH (a:$LABEL1 {$PROP1: $VALUE})-[:$REL1]->(b:$LABEL2) WITH b MATCH (b)-[:$REL2]->(c:$LABEL3) WHERE c.$FILTER_PROP = $FILTER_VAL WITH b, count(c) AS cnt WHERE cnt > $THRESHOLD RETURN b, cnt
\end{lstlisting}

\subsubsection{Anti}
\begin{lstlisting}[style=cypherstyle]
T004:
MATCH (a:$LABEL) WHERE NOT a.$PROP $OP $VALUE RETURN count(a) AS cnt

T006:
MATCH (a:$LABEL1 {$PROP1: $VALUE}), (b:$LABEL2) WHERE NOT (a)-[:$REL_TYPE]->(b) RETURN count(b) AS cnt

T007:
MATCH (a:$LABEL1 {$PROP1: $VALUE}), (b:$LABEL2) WHERE NOT (a)-[:$REL_TYPE]->(b) RETURN a, count(b) AS cnt

T008:
MATCH (a:$LABEL) WHERE NOT (a)-[:$REL1]->(:$LABEL2) OR (a)-[:$REL2]->(:$LABEL3) RETURN count(a) AS cnt
\end{lstlisting}

\subsubsection{Union}
\begin{lstlisting}[style=cypherstyle]
T002:
MATCH (a:$LABEL1)-[:$REL1]->(b:$LABEL2) WITH a, $AGG_FUNC1(b.$RET_PROP1) AS metric1 WHERE metric1 $OP1 $THRESHOLD1 RETURN a.$PROP1 AS EntityName, '$CATEGORY1' AS Category, metric1 AS ValueScore UNION ALL MATCH (c:$LABEL3)<-[:$REL2]-(d:$LABEL4) WITH c, $AGG_FUNC2(d.$RET_PROP3) AS metric2 WHERE metric2 $OP2 $THRESHOLD2 RETURN c.$PROP2 AS EntityName, '$CATEGORY2' AS Category, metric2 AS ValueScore

T003:
MATCH (a:$LABEL1) RETURN count(a.$PROP) AS cnt UNION MATCH (b:$LABEL2) RETURN count(b.$PROP) AS cnt
\end{lstlisting}

\subsubsection{Triadic}
\begin{lstlisting}[style=cypherstyle]
T002:
MATCH (a:$LABEL1)-[:$REL]-()-[:$REL]-(b:$LABEL2) WHERE NOT (a)-[:$REL]-(b) WITH a, count(DISTINCT b) AS cnt RETURN a.$PROP, cnt
\end{lstlisting}

\subsection{Judge Templates}
This section lists the query templates used for judge-oriented query sampling. Each template is paired with an anti-template when available, so that the instantiated query can be used to compare positive and contrastive graph conditions.

\subsubsection{Basic Aggregation}
\begin{lstlisting}[style=cypherstyle]
T002:
MATCH (g:$GROUP_LABEL)<-[:$REL]-(n:$LABEL) WITH g, n RETURN g.$GROUP_PROP AS $GROUP_ALIAS, collect(n.$NODE_PROP)[0..1] AS $COLLECT_ALIAS
Anti:
MATCH (g:$GROUP_LABEL), (n:$LABEL) WHERE NOT (g)<-[:$REL]-(n) WITH g, n RETURN g.$GROUP_PROP AS $GROUP_ALIAS, collect(n.$NODE_PROP)[0..1] AS $COLLECT_ALIAS

T006:
MATCH (a:$LABEL1 {$PROP1: $VALUE}), (b:$LABEL2) WHERE NOT (a)-[:$REL]->(b) WITH b ORDER BY b.$PROP_ID LIMIT 32 RETURN b.$PROP_ID
Anti:
MATCH (a:$LABEL1 {$PROP1: $VALUE}), (b:$LABEL2) WHERE (a)-[:$REL]->(b) WITH b ORDER BY b.$PROP_ID LIMIT 32 RETURN b.$PROP_ID

\end{lstlisting}

\subsubsection{Nested-Loop}
\begin{lstlisting}[style=cypherstyle]
T008:
MATCH (a:$LABEL1 {$PROP1: $VALUE}) MATCH (b:$LABEL2 {$PROP2: a.$REF_PROP}) WITH a, b ORDER BY b.$PROP_ID RETURN a, collect(b.$PROP_ID)[0..1] AS bs
Anti:
MATCH (a:$LABEL1 {$PROP1: $VALUE}) MATCH (b:$LABEL2) WHERE b.$PROP2 <> a.$REF_PROP WITH a, b ORDER BY b.$PROP_ID RETURN a, collect(b.$PROP_ID)[0..1] AS bs

T010:
MATCH (a:$LABEL1) WHERE (a)-[:$REL_TYPE]->(:$LABEL2) WITH a RETURN collect(a.$RETURN_PROP)[0..1] AS values
Anti:
MATCH (a:$LABEL1) WHERE NOT (a)-[:$REL_TYPE]->(:$LABEL2) WITH a RETURN collect(a.$RETURN_PROP)[0..1] AS values

\end{lstlisting}

\subsubsection{Chain}
\begin{lstlisting}[style=cypherstyle]
T007:
MATCH (a:$LABEL1)-[:$REL1]->(b:$LABEL2)-[:$REL2]->(c:$LABEL3)-[:$REL3]->(d:$LABEL4) WITH a, d RETURN a, collect(d.$PROP)[0..1] AS ds
Anti:
MATCH (a:$LABEL1), (d:$LABEL4) WHERE NOT (a)-[:$REL1]->(b:$LABEL2)-[:$REL2]->(c:$LABEL3)-[:$REL3]->(d) WITH a, d RETURN a, collect(d.$PROP)[0..1] AS ds

\end{lstlisting}

\subsubsection{Join}
\begin{lstlisting}[style=cypherstyle]
T002:
MATCH (a:$LABEL1), (b:$LABEL2) WHERE a.$PROP1 = b.$PROP2 AND a <> b WITH a, b RETURN a, b
Anti:
MATCH (a:$LABEL1), (b:$LABEL2) WHERE a.$PROP1 <> b.$PROP2 AND a <> b WITH a, b RETURN a, b

T005:
MATCH (a:$LABEL1), (b:$LABEL2) WHERE a.$PROP1 = b.$PROP2 AND a <> b WITH a, b RETURN a, collect(b.$PROP)[0..1] AS bs
Anti:
MATCH (a:$LABEL1), (b:$LABEL2) WHERE a.$PROP1 <> b.$PROP2 AND a <> b WITH a, b RETURN a, collect(b.$PROP)[0..1] AS bs

\end{lstlisting}

\subsubsection{Anti}
\begin{lstlisting}[style=cypherstyle]
T001:
MATCH (a:$LABEL) WHERE NOT a.$PROP $OP $VALUE WITH a ORDER BY a.$PROP_ID LIMIT 10 RETURN a.$PROP_ID
Anti:
MATCH (a:$LABEL) WHERE a.$PROP $OP $VALUE WITH a ORDER BY a.$PROP_ID LIMIT 10 RETURN a.$PROP_ID

T005:
MATCH (a:$LABEL) WHERE NOT a.$PROP $OP $VALUE WITH a RETURN collect(a.$RET_PROP)[0..1] AS vals
Anti:
MATCH (a:$LABEL) WHERE a.$PROP $OP $VALUE WITH a RETURN collect(a.$RET_PROP)[0..1] AS vals

T009:
MATCH (a:$LABEL) WHERE NOT (a)-[:$REL1]->(:$LABEL2) OR (a)-[:$REL2]->(:$LABEL3) WITH a RETURN collect(a.$RET_PROP)[0..1] AS vals

\end{lstlisting}

\subsubsection{Triadic}
\begin{lstlisting}[style=cypherstyle]
T001:
MATCH (a:$LABEL1)-[:$REL]-()-[:$REL]-(b:$LABEL2) WHERE NOT (a)-[:$REL]-(b) WITH b ORDER BY b.$PROP_ID LIMIT 10 RETURN b.$PROP_ID
Anti:
MATCH (a:$LABEL1)-[:$REL]-()-[:$REL]-(b:$LABEL2) WHERE (a)-[:$REL]-(b) WITH b ORDER BY b.$PROP_ID LIMIT 10 RETURN b.$PROP_ID

\end{lstlisting}

\subsection{Management Templates}
This section lists the management-oriented templates used to instantiate graph update queries. Each template contains a write operation together with pre-validation and post-validation analytical queries, enabling us to test whether graph updates produce the expected observable changes in downstream query results.

\subsubsection{Create Templates}
\begin{lstlisting}[style=cypherstyle]
T001:
Pre-validation:
MATCH (a:$LABEL) RETURN avg(a.$PROP) AS avg_value
Update:
CREATE (n:$LABEL {$PROP: $VALUE})
Post-validation:
MATCH (a:$LABEL) RETURN avg(a.$PROP) AS avg_value

T002:
Pre-validation:
MATCH (g:$GROUP_LABEL)<-[:$REL]-(n:$LABEL) RETURN g.$GROUP_PROP, count(n) AS cnt
Update:
CREATE (n:$LABEL {$PROP: $NAME})-[:$REL]->(g:$GROUP_LABEL {$PROP_ID: $GID})
Post-validation:
MATCH (g:$GROUP_LABEL)<-[:$REL]-(n:$LABEL) RETURN g.$GROUP_PROP, count(n) AS cnt

T003:
Pre-validation:
MATCH (a:$L1)-[:$R1]->(b:$L2)-[:$R2]->(c:$L3) RETURN count(*) AS cnt
Update:
MATCH (a:$L1 {$PROP_ID1: $ID1}), (b:$L2 {$PROP_ID2: $ID2}) CREATE (a)-[:$R1]->(b)
Post-validation:
MATCH (a:$L1)-[:$R1]->(b:$L2)-[:$R2]->(c:$L3) RETURN count(*) AS cnt

T004:
Pre-validation:
MATCH (a:$LABEL1)-[r:$REL_TYPE]-(b:$LABEL2) WHERE r.$REL_PROP > $VAL RETURN count(a) AS cnt
Update:
CREATE (a:$LABEL1)-[r:$REL_TYPE {$REL_PROP: $NEW_VAL}]->(b:$LABEL2 {$PROP_ID: $BID})
Post-validation:
MATCH (a:$LABEL1)-[r:$REL_TYPE]-(b:$LABEL2) WHERE r.$REL_PROP > $VAL RETURN count(a) AS cnt

T005:
Pre-validation:
MATCH (a:$L1) MATCH (b:$L2) WHERE a.$P1 = b.$P2 RETURN count(b) AS cnt
Update:
CREATE (n:$L1 {$P1: $VAL})
Post-validation:
MATCH (a:$L1) MATCH (b:$L2) WHERE a.$P1 = b.$P2 RETURN count(b) AS cnt

T006:
Pre-validation:
MATCH p = (a:$L1)-[:$R *$MIN_HOPS..$MAX_HOPS]-(b:$L2) RETURN avg(length(p)) AS avg_len
Update:
MATCH (a:$L1 {$PROP_ID1: $ID1}), (b:$L2 {$PROP_ID2: $ID2}) CREATE (a)-[:$R]->(b)
Post-validation:
MATCH p = (a:$L1)-[:$R *$MIN_HOPS..$MAX_HOPS]-(b:$L2) RETURN avg(length(p)) AS avg_len

T007:
Pre-validation:
MATCH (g:$GL {$PROP_ID: $GID})<-[:$R]-(n:$L {$P: $V}) RETURN count(n) AS exists
Update:
CREATE (n:$L {$P: $V})-[:$R]->(g:$GL {$PROP_ID: $GID})
Post-validation:
MATCH (g:$GL {$PROP_ID: $GID})<-[:$R]-(n:$L {$P: $V}) RETURN count(n) AS exists

T008:
Pre-validation:
MATCH (a:$L1), (b:$L2) WHERE a.$P1 = b.$P2 AND a <> b RETURN count(*) AS cnt
Update:
CREATE (n:$L1 {$P1: $VAL})
Post-validation:
MATCH (a:$L1), (b:$L2) WHERE a.$P1 = b.$P2 AND a <> b RETURN count(*) AS cnt

T009:
Pre-validation:
MATCH (a:$L1 {$P1: $V}), (b:$L2) WHERE NOT (a)-[:$R]->(b) RETURN count(b) AS cnt
Update:
MATCH (a:$L1 {$PROP_ID1: $AID}), (b:$L2 {$PROP_ID2: $BID}) CREATE (a)-[:$R]->(b)
Post-validation:
MATCH (a:$L1 {$P1: $V}), (b:$L2) WHERE NOT (a)-[:$R]->(b) RETURN count(b) AS cnt

T010:
Pre-validation:
MATCH (a:$L1)-[:$R1]->(b:$L2)-[:$R2]->(c:$L3)-[:$R3]->(d:$L4) RETURN avg(d.$NP) AS final_avg
Update:
CREATE (n:$L4 {$NP: $VAL})<-[:$R3]-(c:$L3 {$PROP_ID: $CID})
Post-validation:
MATCH (a:$L1)-[:$R1]->(b:$L2)-[:$R2]->(c:$L3)-[:$R3]->(d:$L4) RETURN avg(d.$NP) AS final_avg

T011:
Pre-validation:
MATCH (a:$L2) RETURN sum(a.$P2) AS total
Update:
CREATE (n:$L1 {$PROP: $VALUE})
Post-validation:
MATCH (a:$L2) RETURN sum(a.$P2) AS total

T012:
Pre-validation:
MATCH (a:$L3)-[:$R2]->(b:$L4) RETURN count(*) AS cnt
Update:
MATCH (a:$L1 {$PROP_ID1: $ID1}), (b:$L2 {$PROP_ID2: $ID2}) CREATE (a)-[:$R1]->(b)
Post-validation:
MATCH (a:$L3)-[:$R2]->(b:$L4) RETURN count(*) AS cnt

T013:
Pre-validation:
MATCH (a:$L2) RETURN avg(a.$P2) AS avg_val
Update:
CREATE (n:$L1 {$P1: $VAL})
Post-validation:
MATCH (a:$L2) RETURN avg(a.$P2) AS avg_val

\end{lstlisting}

\subsubsection{Merge Templates}
\begin{lstlisting}[style=cypherstyle]
T001:
Pre-validation:
MATCH (a:$LABEL) RETURN sum(a.$PROP) AS total
Update:
MERGE (n:$LABEL {$PROP_ID: $VALUE}) SET n.$PROP = $VAL
Post-validation:
MATCH (a:$LABEL) RETURN sum(a.$PROP) AS total

T002:
Pre-validation:
MATCH (n:$L1 {$P: $V})-[:$R]->(m) RETURN count(m) AS cnt
Update:
MATCH (a:$L1 {$PROP_ID1: $ID1}), (b:$L2 {$PROP_ID2: $ID2}) MERGE (a)-[:$R]->(b)
Post-validation:
MATCH (n:$L1 {$P: $V})-[:$R]->(m) RETURN count(m) AS cnt

T003:
Pre-validation:
MATCH (a:$L1)-[:$R1]->(b:$L2) MATCH (a)-[:$R2]->(b) RETURN count(*) AS cnt
Update:
MATCH (a:$L1 {$PROP_ID1: $ID1}), (b:$L2 {$PROP_ID2: $ID2}) MERGE (a)-[:$R2]->(b)
Post-validation:
MATCH (a:$L1)-[:$R1]->(b:$L2) MATCH (a)-[:$R2]->(b) RETURN count(*) AS cnt

T004:
Pre-validation:
MATCH (a:$L1 {$P1: $V}) MATCH (b:$L2 {$P2: a.$P1}) RETURN count(b) AS cnt
Update:
MERGE (n:$L1 {$PROP_ID: $VALUE}) SET n.$P1 = $NEW_REF
Post-validation:
MATCH (a:$L1 {$P1: $V}) MATCH (b:$L2 {$P2: a.$P1}) RETURN count(b) AS cnt

T005:
Pre-validation:
MATCH (n:$L) WHERE (n)-[:$R1]->(:$L1) OR (n)-[:$R2]->(:$L2) RETURN count(n) AS cnt
Update:
MATCH (n:$L {$PROP_ID: $VALUE}), (m:$L2 {$PROP_ID: $MID}) MERGE (n)-[:$R2]->(m)
Post-validation:
MATCH (n:$L) WHERE (n)-[:$R1]->(:$L1) OR (n)-[:$R2]->(:$L2) RETURN count(n) AS cnt

T006:
Pre-validation:
MATCH (a:$L1)-[:$R1]->(b:$L2)-[:$R2]->(c:$L3) RETURN avg(c.$P) AS avg_val
Update:
MERGE (n:$L3 {$PROP_ID: $VALUE}) SET n.$P = $VAL
Post-validation:
MATCH (a:$L1)-[:$R1]->(b:$L2)-[:$R2]->(c:$L3) RETURN avg(c.$P) AS avg_val

T007:
Pre-validation:
MATCH (a:$L1) CALL {WITH a MATCH (b:$L2) WHERE a.$P1 = b.$P2 RETURN b.$BP AS val ORDER BY b.$BP DESC LIMIT 10} RETURN a, val
Update:
MERGE (n:$L2 {$PROP_ID: $VALUE}) SET n.$BP = $NEW_VAL
Post-validation:
MATCH (a:$L1) CALL {WITH a MATCH (b:$L2) WHERE a.$P1 = b.$P2 RETURN b.$BP AS val ORDER BY b.$BP DESC LIMIT 10} RETURN a, val

T008:
Pre-validation:
MATCH (a:$L1)-[:$R *$MIN_HOPS..$MAX_HOPS]-(b:$L2) RETURN count(b) AS cnt
Update:
MATCH (a:$L1 {$PROP_ID1: $ID1}), (b:$L2 {$PROP_ID2: $ID2}) MERGE (a)-[:$R]->(b)
Post-validation:
MATCH (a:$L1)-[:$R *$MIN_HOPS..$MAX_HOPS]-(b:$L2) RETURN count(b) AS cnt

T009:
Pre-validation:
MATCH (a:$L1), (b:$L2) WHERE a.$P1 = b.$P2 WITH a, count(b) AS cnt WHERE cnt > $THRESHOLD RETURN a
Update:
MERGE (n:$L2 {$PROP_ID: $VALUE, $P2: $REF_VAL})
Post-validation:
MATCH (a:$L1), (b:$L2) WHERE a.$P1 = b.$P2 WITH a, count(b) AS cnt WHERE cnt > $THRESHOLD RETURN a

T010:
Pre-validation:
MATCH (s:$SL {$SP: $SV}) MATCH p = (s) ((n1:$L1)<-[:$R1]-(r:$RL)-[:$R2]->(n2:$L2) WHERE r.$NP = $NV){$MIN_HOPS..$MAX_HOPS} (e:$EL) RETURN count(p) AS path_cnt
Update:
MERGE (r:$RL {$PROP_ID: $VALUE}) SET r.$NP = $NV
Post-validation:
MATCH (s:$SL {$SP: $SV}) MATCH p = (s) ((n1:$L1)<-[:$R1]-(r:$RL)-[:$R2]->(n2:$L2) WHERE r.$NP = $NV){$MIN_HOPS..$MAX_HOPS} (e:$EL) RETURN count(p) AS path_cnt

T011:
Pre-validation:
MATCH (a:$L2) RETURN max(a.$P2) AS max_val
Update:
MERGE (n:$L1 {$PROP_ID: $VALUE}) SET n.$P1 = $VAL
Post-validation:
MATCH (a:$L2) RETURN max(a.$P2) AS max_val

T012:
Pre-validation:
MATCH p = (a:$L3)-[:$R2 *$MIN_HOPS..$MAX_HOPS]-(b:$L4) RETURN avg(length(p)) AS avg_len
Update:
MATCH (a:$L1 {$PROP_ID1: $ID1}), (b:$L2 {$PROP_ID2: $ID2}) MERGE (a)-[:$R1]->(b)
Post-validation:
MATCH p = (a:$L3)-[:$R2 *$MIN_HOPS..$MAX_HOPS]-(b:$L4) RETURN avg(length(p)) AS avg_len

T013:
Pre-validation:
MATCH (a:$L3), (b:$L4) WHERE a.$P3 = b.$P4 RETURN count(*) AS cnt
Update:
MERGE (n:$L1 {$PROP_ID: $VALUE}) SET n.$P1 = $VAL
Post-validation:
MATCH (a:$L3), (b:$L4) WHERE a.$P3 = b.$P4 RETURN count(*) AS cnt

\end{lstlisting}

\subsubsection{Delete Templates}
\begin{lstlisting}[style=cypherstyle]
T001:
Pre-validation:
MATCH (a:$LABEL) RETURN max(a.$PROP) AS max_val
Update:
MATCH (n:$LABEL {$PROP_ID: $VALUE}) DELETE n
Post-validation:
MATCH (a:$LABEL) RETURN max(a.$PROP) AS max_val

T002:
Pre-validation:
MATCH (n:$L1 {$P: $V})<-[:$R]-(m) RETURN count(m) AS cnt
Update:
MATCH (a)-[r:$R]->(b:$L1 {$PROP_ID: $VALUE}) DELETE r
Post-validation:
MATCH (n:$L1 {$P: $V})<-[:$R]-(m) RETURN count(m) AS cnt

T003:
Pre-validation:
MATCH (a:$L1), (b:$L2) WHERE a.$P1 = b.$P2 RETURN count(*) AS cnt
Update:
MATCH (n:$L2 {$PROP_ID: $VALUE}) DETACH DELETE n
Post-validation:
MATCH (a:$L1), (b:$L2) WHERE a.$P1 = b.$P2 RETURN count(*) AS cnt

T004:
Pre-validation:
MATCH (a:$L1 {$P1: $V}), (b:$L2) WHERE NOT (a)-[:$R]->(b) RETURN count(b) AS cnt
Update:
MATCH (a:$L1 {$PROP_ID1: $AID})-[r:$R]->(b:$L2 {$PROP_ID2: $BID}) DELETE r
Post-validation:
MATCH (a:$L1 {$P1: $V}), (b:$L2) WHERE NOT (a)-[:$R]->(b) RETURN count(b) AS cnt

T005:
Pre-validation:
MATCH (a:$L1)-[:$R1]->(b:$L2)-[:$R2]->(c:$L3) RETURN count(c) AS cnt
Update:
MATCH (b:$L2 {$PROP_ID: $VALUE}) DETACH DELETE b
Post-validation:
MATCH (a:$L1)-[:$R1]->(b:$L2)-[:$R2]->(c:$L3) RETURN count(c) AS cnt

T006:
Pre-validation:
MATCH (n:$L) WHERE (n)-[:$R1]->(:$L1) OR (n)-[:$R2]->(:$L2) RETURN avg(n.$NP)
Update:
MATCH (n:$L {$PROP_ID: $VALUE}) DETACH DELETE n
Post-validation:
MATCH (n:$L) WHERE (n)-[:$R1]->(:$L1) OR (n)-[:$R2]->(:$L2) RETURN avg(n.$NP)

T007:
Pre-validation:
MATCH p = (a:$L1)-[:$R *$MIN_HOPS..$MAX_HOPS]-(b:$L2) RETURN avg(length(p)) AS avg_len
Update:
MATCH (n {$PROP_ID: $VALUE}) DETACH DELETE n
Post-validation:
MATCH p = (a:$L1)-[:$R *$MIN_HOPS..$MAX_HOPS]-(b:$L2) RETURN avg(length(p)) AS avg_len

T008:
Pre-validation:
MATCH (target:$L {$PROP_ID: $VALUE})-[:$R]->(g:$GL) WITH g MATCH (g)<-[:$R]-(n:$L) WITH g, n ORDER BY n.$P DESC LIMIT 3 RETURN g.$GP, n.$P AS top_val
Update:
MATCH (n:$L {$PROP_ID: $VALUE}) DETACH DELETE n
Post-validation:
MATCH (g:$GL {$GP: $PREVIOUS_GP_VALUE}) MATCH (g)<-[:$R]-(n:$L) WITH g, n ORDER BY n.$P DESC LIMIT 3 RETURN g.$GP, n.$P AS top_val

T009:
Pre-validation:
MATCH (a:$L1), (b:$L2) WHERE a.$P1 = b.$P2 WITH a, count(b) AS cnt WHERE cnt > $THRESHOLD RETURN a
Update:
MATCH (n:$L2 {$PROP_ID: $VALUE}) DETACH DELETE n
Post-validation:
MATCH (a:$L1), (b:$L2) WHERE a.$P1 = b.$P2 WITH a, count(b) AS cnt WHERE cnt > $THRESHOLD RETURN a

T010:
Pre-validation:
MATCH (start:$SL {$SP: $SV}) MATCH p = (start) (...){$MIN_HOPS..$MAX_HOPS} (end:$EL) RETURN count(p) AS cnt
Update:
MATCH (n {$PROP_ID: $VALUE}) DETACH DELETE n
Post-validation:
MATCH (start:$SL {$SP: $SV}) MATCH p = (start) (...){$MIN_HOPS..$MAX_HOPS} (end:$EL) RETURN count(p) AS cnt

T011:
Pre-validation:
MATCH (a:$L2) RETURN min(a.$P2) AS min_val
Update:
MATCH (n:$L1 {$PROP_ID: $VALUE}) DETACH DELETE n
Post-validation:
MATCH (a:$L2) RETURN min(a.$P2) AS min_val

T012:
Pre-validation:
MATCH (a:$L3)-[:$R2]->(b:$L4) RETURN count(*) AS cnt
Update:
MATCH (a)-[r:$R1]->(b:$L1 {$PROP_ID: $VALUE}) DELETE r
Post-validation:
MATCH (a:$L3)-[:$R2]->(b:$L4) RETURN count(*) AS cnt

T013:
Pre-validation:
MATCH (a:$L2) RETURN avg(a.$P2) AS avg_val, count(a) AS cnt
Update:
MATCH (n:$L1 {$PROP_ID: $VALUE}) DETACH DELETE n
Post-validation:
MATCH (a:$L2) RETURN avg(a.$P2) AS avg_val, count(a) AS cnt

\end{lstlisting}

\subsubsection{Set Templates}
\begin{lstlisting}[style=cypherstyle]
T001:
Pre-validation:
MATCH (a:$LABEL) RETURN avg(a.$PROP) AS avg_val
Update:
MATCH (n:$LABEL {$PROP_ID: $VALUE}) SET n.$PROP = $VAL
Post-validation:
MATCH (a:$LABEL) RETURN avg(a.$PROP) AS avg_val

T002:
Pre-validation:
MATCH (a:$L1)-[r:$R]-(b:$L2) WHERE r.$RP $OP $V RETURN count(*) AS cnt
Update:
MATCH (a {$PROP_ID1: $AID})-[r:$R]->(b {$PROP_ID2: $BID}) SET r.$RP = $VALUE
Post-validation:
MATCH (a:$L1)-[r:$R]-(b:$L2) WHERE r.$RP $OP $V RETURN count(*) AS cnt

T003:
Pre-validation:
MATCH (a:$L1), (b:$L2) WHERE a.$P1 = b.$P2 RETURN count(*) AS cnt
Update:
MATCH (n:$L1 {$PROP_ID: $VALUE}) SET n.$P1 = $NEW_VAL
Post-validation:
MATCH (a:$L1), (b:$L2) WHERE a.$P1 = b.$P2 RETURN count(*) AS cnt

T004:
Pre-validation:
MATCH (a:$L1)-[:$R1]->(b:$L2)-[:$R2]->(c:$L3) RETURN sum(c.$P) AS total
Update:
MATCH (n:$L3 {$PROP_ID: $VALUE}) SET n.$P = $VAL
Post-validation:
MATCH (a:$L1)-[:$R1]->(b:$L2)-[:$R2]->(c:$L3) RETURN sum(c.$P) AS total

T005:
Pre-validation:
MATCH (a:$L1) MATCH (b:$L2 {$P2: a.$P1}) RETURN count(b) AS cnt
Update:
MATCH (n:$L1 {$PROP_ID: $VALUE}) SET n.$P1 = $NEW_REF
Post-validation:
MATCH (a:$L1) MATCH (b:$L2 {$P2: a.$P1}) RETURN count(b) AS cnt

T006:
Pre-validation:
MATCH (n:$L) WHERE n.$P $OP $V OR (n)-[:$R]->(:$L2) RETURN count(n) AS cnt
Update:
MATCH (n:$L {$PROP_ID: $VALUE}) SET n.$P = $NEW_V
Post-validation:
MATCH (n:$L) WHERE n.$P $OP $V OR (n)-[:$R]->(:$L2) RETURN count(n) AS cnt

T007:
Pre-validation:
MATCH (n:$L {$PROP_ID: $VALUE}) RETURN n.$P AS old_value
Update:
MATCH (n:$L {$PROP_ID: $VALUE}) SET n.$P = $VAL
Post-validation:
MATCH (n:$L {$PROP_ID: $VALUE}) RETURN n.$P AS new_value

T008:
Pre-validation:
MATCH (a:$L1)-[:$R *$MIN_HOPS..$MAX_HOPS]-(b:$L2) RETURN count(b) AS cnt
Update:
MATCH (n {$PROP_ID: $VALUE}) SET n:$L2
Post-validation:
MATCH (a:$L1)-[:$R *$MIN_HOPS..$MAX_HOPS]-(b:$L2) RETURN count(b) AS cnt

T009:
Pre-validation:
MATCH (a:$L1), (b:$L2) WHERE a.$P1 = b.$P2 WITH a, avg(b.$BP) AS avg_b WHERE avg_b > $THRESHOLD RETURN a
Update:
MATCH (n:$L2 {$PROP_ID: $VALUE}) SET n.$BP = $VAL
Post-validation:
MATCH (a:$L1), (b:$L2) WHERE a.$P1 = b.$P2 WITH a, avg(b.$BP) AS avg_b WHERE avg_b > $THRESHOLD RETURN a

T010:
Pre-validation:
MATCH p = (a:$L1)-[:$R *$MIN_HOPS..$MAX_HOPS]-(b:$L2) RETURN avg(length(p)) AS avg_len
Update:
MATCH (a {$PROP_ID1: $AID}), (b {$PROP_ID2: $BID}) CREATE (a)-[:$R]->(b)
Post-validation:
MATCH p = (a:$L1)-[:$R *$MIN_HOPS..$MAX_HOPS]-(b:$L2) RETURN avg(length(p)) AS avg_len

T011:
Pre-validation:
MATCH (a:$L2) RETURN sum(a.$P2) AS total
Update:
MATCH (n:$L1 {$PROP_ID: $VALUE}) SET n.$P1 = $VAL
Post-validation:
MATCH (a:$L2) RETURN sum(a.$P2) AS total

T012:
Pre-validation:
MATCH (a:$L3)-[r:$R2]-(b:$L4) WHERE r.$RP $OP $V RETURN count(*) AS cnt
Update:
MATCH (a {$PROP_ID1: $AID})-[r:$R1]->(b {$PROP_ID2: $BID}) SET r.$RP = $VALUE
Post-validation:
MATCH (a:$L3)-[r:$R2]-(b:$L4) WHERE r.$RP $OP $V RETURN count(*) AS cnt

T013:
Pre-validation:
MATCH (a:$L3)-[:$R2]->(b:$L4)-[:$R3]->(c:$L5) RETURN count(*) AS cnt
Update:
MATCH (n {$PROP_ID: $VALUE}) SET n:$L2
Post-validation:
MATCH (a:$L3)-[:$R2]->(b:$L4)-[:$R3]->(c:$L5) RETURN count(*) AS cnt

\end{lstlisting}

\subsubsection{Foreach Templates}
\begin{lstlisting}[style=cypherstyle]
T001:
Pre-validation:
MATCH (n:$LABEL) RETURN count(n) AS cnt
Update:
FOREACH (x IN $LIST | CREATE (:$LABEL {val: x}))
Post-validation:
MATCH (n:$LABEL) RETURN count(n) AS cnt

T002:
Pre-validation:
MATCH (n:$L1 {$P: $V})-[:$R]->(m) RETURN count(m) AS cnt
Update:
MATCH (a:$L1 {$PROP_ID1: $AID}), (others:$L2) WHERE others.$PROP_ID2 IN $IDS FOREACH (o IN [others] | CREATE (a)-[:$R]->(o))
Post-validation:
MATCH (n:$L1 {$P: $V})-[:$R]->(m) RETURN count(m) AS cnt

T003:
Pre-validation:
MATCH (a:$L1), (b:$L2) WHERE a.$P1 = b.$P2 RETURN count(*) AS cnt
Update:
MATCH (n:$L1) WHERE n.$PROP_ID IN $IDS FOREACH (i IN [n] | SET i.$P1 = $NEW_VAL)
Post-validation:
MATCH (a:$L1), (b:$L2) WHERE a.$P1 = b.$P2 RETURN count(*) AS cnt

T004:
Pre-validation:
MATCH (a:$L1)-[:$R1]->(b:$L2)-[:$R2]->(c:$L3) RETURN avg(c.$P) AS avg_val
Update:
MATCH (n:$L3) WHERE n.$PROP_ID IN $IDS FOREACH (i IN [n] | SET i.$P = $VAL)
Post-validation:
MATCH (a:$L1)-[:$R1]->(b:$L2)-[:$R2]->(c:$L3) RETURN avg(c.$P) AS avg_val

T005:
Pre-validation:
MATCH (a:$L1) MATCH (b:$L2 {$P2: a.$P1}) RETURN count(b) AS cnt
Update:
MATCH (n:$L1) WHERE n.$PROP_ID IN $IDS FOREACH (i IN [n] | SET i.$P1 = $NEW_REF)
Post-validation:
MATCH (a:$L1) MATCH (b:$L2 {$P2: a.$P1}) RETURN count(b) AS cnt

T006:
Pre-validation:
MATCH (n:$L) WHERE (n)-[:$R1]->(:$L1) OR (n)-[:$R2]->(:$L2) RETURN sum(n.$NP)
Update:
MATCH (n:$L) WHERE n.$PROP_ID IN $IDS FOREACH (i IN [n] | CREATE (i)-[:$R1]->(:$L1))
Post-validation:
MATCH (n:$L) WHERE (n)-[:$R1]->(:$L1) OR (n)-[:$R2]->(:$L2) RETURN sum(n.$NP)

T007:
Pre-validation:
MATCH (n:$L) WHERE n.$PROP_ID IN $IDS RETURN n.$PROP_ID AS id, n.$P AS val ORDER BY id
Update:
MATCH (n:$L) WHERE n.$PROP_ID IN $IDS FOREACH (i IN [n] | SET i.$P = i.$P + $INC)
Post-validation:
MATCH (n:$L) WHERE n.$PROP_ID IN $IDS RETURN n.$PROP_ID AS id, n.$P AS val ORDER BY id

T008:
Pre-validation:
MATCH p = (a:$L1)-[:$R *$MIN_HOPS..$MAX_HOPS]-(b:$L2) RETURN avg(length(p)) AS avg_len
Update:
MATCH (a:$L1), (b:$L2) WHERE a.$PROP_ID IN $A_IDS AND b.$PROP_ID IN $B_IDS FOREACH (x IN [1] | CREATE (a)-[:$R]->(b))
Post-validation:
MATCH p = (a:$L1)-[:$R *$MIN_HOPS..$MAX_HOPS]-(b:$L2) RETURN avg(length(p)) AS avg_len

T009:
Pre-validation:
MATCH (a:$L1), (b:$L2) WHERE a.$P1 = b.$P2 WITH a, count(b) AS cnt WHERE cnt > $THRESHOLD RETURN a
Update:
MATCH (n:$L2) WHERE n.$PROP_ID IN $IDS FOREACH (i IN [n] | SET i.$P2 = $NEW_REF)
Post-validation:
MATCH (a:$L1), (b:$L2) WHERE a.$P1 = b.$P2 WITH a, count(b) AS cnt WHERE cnt > $THRESHOLD RETURN a

T010:
Pre-validation:
MATCH (s:$SL {$SP: $SV}) MATCH p = (s) ((n1:$L1)<-[:$R1]-(r:$RL)-[:$R2]->(n2:$L2) WHERE r.$NP = $NV){$MIN_HOPS..$MAX_HOPS} (e:$EL) RETURN count(p) AS cnt
Update:
MATCH (r:$RL) WHERE r.$PROP_ID IN $IDS FOREACH (i IN [r] | SET i.$NP = $NV)
Post-validation:
MATCH (s:$SL {$SP: $SV}) MATCH p = (s) ((n1:$L1)<-[:$R1]-(r:$RL)-[:$R2]->(n2:$L2) WHERE r.$NP = $NV){$MIN_HOPS..$MAX_HOPS} (e:$EL) RETURN count(p) AS cnt

T011:
Pre-validation:
MATCH (a:$L2) RETURN avg(a.$P2) AS avg_val, count(a) AS cnt
Update:
MATCH (n:$L1) WHERE n.$PROP_ID IN $IDS FOREACH (i IN [n] | SET i.$P1 = $VAL)
Post-validation:
MATCH (a:$L2) RETURN avg(a.$P2) AS avg_val, count(a) AS cnt

T012:
Pre-validation:
MATCH (a:$L3)-[:$R2]->(b:$L4) RETURN count(*) AS cnt
Update:
MATCH (a:$L1), (b:$L2) WHERE a.$PROP_ID IN $A_IDS AND b.$PROP_ID IN $B_IDS FOREACH (x IN [1] | CREATE (a)-[:$R1]->(b))
Post-validation:
MATCH (a:$L3)-[:$R2]->(b:$L4) RETURN count(*) AS cnt

T013:
Pre-validation:
MATCH (a:$L3), (b:$L4) WHERE a.$P3 = b.$P4 RETURN count(*) AS cnt
Update:
MATCH (n:$L1) WHERE n.$PROP_ID IN $IDS FOREACH (i IN [n] | SET i.$P1 = $NEW_VAL)
Post-validation:
MATCH (a:$L3), (b:$L4) WHERE a.$P3 = b.$P4 RETURN count(*) AS cnt

\end{lstlisting}

\section{Detailed Evaluation Metrics}
\label{app:metrics}

In this section, we provide detailed definitions and mathematical formulations for the evaluation metrics used in our experiments. The metrics are categorized based on the output characteristics of the three distinct experimental settings: \emph{No Aggregation (NoAgg)}, \emph{Boolean Queries}, and \emph{Aggregation (Agg)}.

\subsection{Set-based Metrics}
In the \emph{NoAgg} setting, the model output and the ground truth are treated as sets of entities. Let $y$ denote the ground-truth set of entities and $\hat{y}$ denote the predicted set of entities for a given query.

\paragraph{Jaccard Similarity.} 
To measure the overall overlap between the predicted set and the ground-truth set, we utilize the Jaccard Similarity coefficient. It is defined as the size of the intersection divided by the size of the union of the two sets:
\begin{equation}
    J(y, \hat{y}) = \frac{|y \cap \hat{y}|}{|y \cup \hat{y}|}.
\end{equation}
A value of 1 indicates a perfect match, while 0 indicates no overlap.

\paragraph{F1 Score.} 
To jointly assess the precision and recall of the retrieved entities, we employ the F1 score. First, Precision ($P$) and Recall ($R$) are calculated as follows:
\begin{equation}
    P = \frac{|y \cap \hat{y}|}{|\hat{y}|}, \quad R = \frac{|y \cap \hat{y}|}{|y|}.
\end{equation}
The F1 score is the harmonic mean of Precision and Recall:
\begin{equation}
    F_1 = 2 \cdot \frac{P \cdot R}{P + R}.
\end{equation}

\subsection{Classification Metrics}

\paragraph{Accuracy.} 
We treat this as a binary classification problem. Let $N$ be the total number of candidate pairs evaluated. Let $y_i \in \{0, 1\}$ be the ground truth label for the $i$-th pair (where 1 represents a valid result pair and 0 otherwise), and let $\hat{y}_i \in \{0, 1\}$ be the model's prediction. Accuracy is defined as:
\begin{equation}
    \text{Accuracy} = \frac{1}{N} \sum_{i=1}^{N} \mathbb{I}(\hat{y}_i = y_i),
\end{equation}
where $\mathbb{I}(\cdot)$ is the indicator function which equals 1 if the condition is true and 0 otherwise.

\begin{table*}[t]
\centering
\small
\setlength{\tabcolsep}{6pt}
\caption{Performance comparison on aggregation queries, where execution on the noisy graph is equivalent to the clean graph.}
\begin{tabular}{l
ccc
ccc
ccc}
\toprule
\textbf{Baseline} &
\multicolumn{3}{c}{\textbf{NGD-Fin}} &
\multicolumn{3}{c}{\textbf{NGD-BI}} &
\multicolumn{3}{c}{\textbf{NGD-Prime}} \\
\cmidrule(lr){2-4} \cmidrule(lr){5-7} \cmidrule(lr){8-10}
&
\textbf{MdRE} & \textbf{MSLE} & \textbf{sMAPE} &
\textbf{MdRE} & \textbf{MSLE} & \textbf{sMAPE} &
\textbf{MdRE} & \textbf{MSLE} & \textbf{sMAPE} \\
\midrule
Neo4j-Text2Cypher& 0.000 &0.055 &0.021  &0.000  &0.000  & 0.001 &0.965 &9.495  & 1.155 \\
GPT5.1-Codex  &  0.000 &  0.001 & 0.001 & 0.000 &3.060  & 0.615 &0.000 &11.476  & 0.541 \\
DeepseekV3.2 &  0.000 &  0.001 & 0.003 & 0.000 & 5.148 & 0.763 & 0.000  &  8.115 & 0.614  \\
Qwen3-Coder  & 0.000 & 0.048 & 0.018 & 0.000 & 2.801 &  0.756 & 0.000& 7.623 & 0.639 \\
GraphRAG           & 0.000 & 0.588 & 0.904 & —— & 0.480 & 1.998 & 0.875 & 2.400 & 1.201 \\
\bottomrule
\end{tabular}
\label{tab:clean}
\end{table*}

\begin{table*}[t]
\centering
\small
\setlength{\tabcolsep}{6pt}
\caption{Performance comparison on aggregation queries, where execution on the noisy graph is inequivalent to the clean graph.}
\begin{tabular}{l
ccc
ccc
ccc}
\toprule
\textbf{Baseline} &
\multicolumn{3}{c}{\textbf{NGD-Fin}} &
\multicolumn{3}{c}{\textbf{NGD-BI}} &
\multicolumn{3}{c}{\textbf{NGD-Prime}} \\
\cmidrule(lr){2-4} \cmidrule(lr){5-7} \cmidrule(lr){8-10}
&
\textbf{MdRE} & \textbf{MSLE} & \textbf{sMAPE} &
\textbf{MdRE} & \textbf{MSLE} & \textbf{sMAPE} &
\textbf{MdRE} & \textbf{MSLE} & \textbf{sMAPE} \\
\midrule
Neo4j-Text2Cypher& 0.179 &27.846 &0.810  & —— & 16.082 &16.509  &——  &46.272  & 1.541 \\
GPT5.1-Codex  &  0.131& 14.940  & 0.771 &0.992 &19.820  &1.233 &0.237 &12.341  & 0.766 \\
DeepseekV3.2 & 0.113 &7.629& 0.451 & 0.995& 25.755 & 1.312 & 0.582  &  13.345 & 0.875  \\
Qwen3-Coder  & 0.116 & 8.231 & 0.446 & 1.000 & 28.878 &  1.337 & 0.578 & 14.185 & 0.891 \\
GraphRAG           & 254.333 & 36.515 & 1.777 & —— & 61.688 & 1.979 & 993.563 & 47.204 & 1.814\\
\bottomrule
\end{tabular}
\label{tab:noisy}
\end{table*}

\begin{table*}[t]
\centering
\scriptsize
\setlength{\tabcolsep}{2.5pt}
\renewcommand{\arraystretch}{0.95}
\caption{Performance comparison of RAG methods on aggregation queries. 
Exact measures exact-match accuracy, $\pm$1 measures the proportion of predictions within one unit of the gold answer, and Null denotes invalid or empty outputs; RMSE, MdRE, MSLE, and sMAPE measure numerical errors. Higher is better for Exact and $\pm$1, while lower is better for Null and all error metrics.}
\resizebox{\textwidth}{!}{
\begin{tabular}{l
rrrrrrr
rrrrrrr}
\toprule
\textbf{Baseline} &
\multicolumn{7}{c}{\textbf{NGD-MCP}} &
\multicolumn{7}{c}{\textbf{NGD-Econ}} \\
\cmidrule(lr){2-8} \cmidrule(lr){9-15}
&
\textbf{Ex.} & \textbf{$\pm$1} & \textbf{Null} & \textbf{RMSE} & \textbf{MdRE} & \textbf{MSLE} & \textbf{sMAPE} &
\textbf{Ex.} & \textbf{$\pm$1} & \textbf{Null} & \textbf{RMSE} & \textbf{MdRE} & \textbf{MSLE} & \textbf{sMAPE} \\
\midrule
HippoRAG2 & 0.088 & 0.250 & 0.619 & 53.012 & 1.000 & 1.051 & 1.196 & 0.012 & 0.038 & 0.097 & 158.308 & 1.000 & 2.138 & 0.926 \\
GraphRAG  & 0.493 & 0.624 & 0.047 & 1.795 & 0.000 & 0.170 & 0.396 & 0.241 & 0.458 & 0.154 & 363623.864 & 0.667 & 0.673 & 0.637 \\
TOG       & 0.266 & 0.342 & 0.044 & 16.169 & 0.000 & 0.359 & 0.380 & 0.260 & 0.323 & 0.125 & 18730.315 & 0.000 & 3.809 & 0.371 \\
\bottomrule
\end{tabular}
}
\label{tab:rag-agg-metrics}
\end{table*}

\subsection{Numerical Error Metrics }
In the \emph{Agg} setting, the output is a scalar numerical value (e.g., from \texttt{COUNT}, \texttt{MAX}, \texttt{MIN}). Let $y_i$ be the ground truth numerical value and $\hat{y}_i$ be the predicted value for the $i$-th query, with a total of $N$ queries.

\paragraph{Median Relative Error (MdRE).} 
To capture the typical prediction error while remaining robust to outliers, we calculate the median of the absolute relative errors:
\begin{equation}
    \text{MdRE} = \text{median}\left( \left\{ \frac{|\hat{y}_i - y_i|}{y_i} \right\}_{i=1}^{N} \right).
\end{equation}
Note that for stability, a small $\epsilon$ is typically added to the denominator if $y_i$ can be zero.

\paragraph{Mean Squared Logarithmic Error (MSLE).} 
To penalize multiplicative deviations rather than additive ones (which is crucial for data with large variances), we use MSLE:
\begin{equation}
    \text{MSLE} = \frac{1}{N} \sum_{i=1}^{N} \left( \log(1 + \hat{y}_i) - \log(1 + y_i) \right)^2.
\end{equation}

\paragraph{Symmetric Mean Absolute Percentage Error (sMAPE).} 
We adopt sMAPE to provide a scale-invariant measure of error that is robust to zero values. It is calculated as the average of the absolute difference divided by the sum of absolute values:
\begin{equation}
    \text{sMAPE} = \frac{100\%}{N} \sum_{i=1}^{N} \frac{|\hat{y}_i - y_i|}{(|\hat{y}_i| + |y_i|) / 2}.
\end{equation}
This metric ranges from 0\% to 200\%, providing a symmetric view of overestimation and underestimation.

\paragraph{Mean Log Relative Error (MLRE).}
We employ MLRE to assess the relative deviation between predicted and actual values, which is particularly useful for capturing order-of-magnitude errors. It is calculated as the mean absolute logarithm of the ratio between the prediction and the ground truth, stabilized by a small constant $\epsilon$:
\begin{equation}
\text{MLRE} = \frac{1}{N} \sum_{i=1}^{N} \left| \ln \left( \frac{\hat{y}_i + \epsilon}{y_i + \epsilon} \right) \right|,
\end{equation}
where $\epsilon$ is a small term (e.g., $10^{-6}$) added to prevent numerical instability. This metric treats the error as a ratio, meaning that a prediction of twice the ground truth yields the same error as a prediction of half the ground truth.

\section{More Experiments}
\label{app:moreexp}
We show more results about aggregation queries on structured domains in Table \ref{tab:clean} and Table \ref{tab:noisy}.

And we show full metrics about aggregation queries on graphs constructed from structured domains in Tabel \ref{tab:rag-agg-metrics}.
\section{Experiment Details}
\subsection{General Failure Case Study}
We summarize representative failure cases of different baselines. These cases illustrate that existing methods fail for different reasons: Text-to-Cypher methods are limited by exact symbolic execution over noisy graphs and may also misunderstand schema semantics, while RAG-style methods suffer from top-$k$ retrieval truncation and fragmented structural evidence. Together, these examples explain why current approaches struggle with robust query answering under noisy, incomplete, and structurally complex data.

\paragraph{Neo4j Text2Cypher.}
We observe two typical failure modes.

\textbf{Case 1: correct query generation but wrong answer due to graph noise.}
For the query ``Find the count of drugs that do not have an \texttt{exposure\_molfunc} relationship with Fluocinonide'', the generated Cypher query is valid. However, executing it on the noisy graph returns \textbf{8807}, while the clean answer is \textbf{7957}. This happens because the method relies on traditional symbolic execution and therefore cannot recover from incomplete-edge noise.

\textbf{Case 2: incorrect query generation due to schema misunderstanding.}
For the query ``Find the count of diseases not associated with Cotinine through a \texttt{molfunc\_molfunc} relationship'', the system incorrectly treats Cotinine as a \texttt{molfunc} entity and produces an invalid Cypher query, leading to the wrong answer \textbf{0}. This shows that Text-to-Cypher baselines may fail both because of graph noise and because of schema or type misunderstanding.

\paragraph{Advanced LLM-based Text2Cypher.}
Stronger LLMs, such as GPT-5.1-Coder, usually avoid obvious query-generation errors, but they still fail when the graph itself is noisy. For example, for the query ``Find account owned by person with create time `2020-01-12 20:52:02.831' and return its node ID'', GPT-5.1-Coder generates the correct Cypher query. However, it returns only one account instead of all three correct accounts, because some ownership edges are missing in the noisy graph. This indicates that even advanced LLM-based Text-to-Cypher methods remain limited by exact execution over incomplete graphs.

\paragraph{RAG-style Methods.}
We observe two main failure modes.

\textbf{Case 1: poor aggregation due to top-$k$ retrieval truncation.}
For the query ``Find the average number of likes across posts'', the correct answer is \textbf{23.7056}, but GraphRAG outputs \textbf{Zero}. The reason is that aggregation requires broad or even complete graph coverage, while retrieval only provides a top-$k$ subset of context, which is insufficient for exact computation.

\textbf{Case 2: failure on multi-hop reasoning because retrieval breaks structural dependencies.}
For the query asking for entities that promote active transportation within 1 to 5 steps, the correct answers include \textit{policies} and \textit{government}. However, GraphRAG returns semantically related but incorrect entities such as \textit{Cities}, \textit{Kenya}, and \textit{Many cities in Europe}. This suggests that retrieval provides fragmented local evidence and does not preserve the full path structure required for accurate multi-hop reasoning.

\subsection{Case Study of RAG-based Methods}
\label{case_study}

We summarize representative success and failure cases of RAG-based methods. These cases show that retrieval-based methods can succeed when the answer is contained in a small set of directly retrieved triples, but often fail when the query requires complete structural evidence, multi-hop path preservation, or exact aggregation. For readability, we abbreviate the retrieved contexts and only show representative snippets.

\paragraph{GraphRAG failure on multi-hop chain queries.}

\begin{itemize}
    \item \textbf{Query:} For the windows performance report gathering task, find the entity representing \texttt{complete\_report} and return its ID along with the IDs of all entities reachable from it within 1 to 5 \texttt{followed\_by} hops.

    \item \textbf{Ground truth:} \texttt{additional metrics}, \texttt{system uptime}, \texttt{network config}, and \texttt{system specs}.

    \item \textbf{Retrieved context:} GraphRAG retrieves several report-related triples, including \path{windows_performance_report_gathering_additional_metrics}
\texttt{ followed\_by }
\path{windows_performance_report_gathering_complete_report},
    together with other partially related task and report triples. However, the retrieved context does not preserve the complete \texttt{followed\_by} chain.

    \item \textbf{Model output:} The model identifies\path{windows_performance_report_gathering_complete_report}, but incorrectly concludes that there are no outgoing \texttt{followed\_by} edges and returns only this entity.

    \item \textbf{Analysis:} This is a typical multi-hop failure. Retrieval provides fragmented local evidence, while the query requires complete path-level structure.
\end{itemize}

\paragraph{GraphRAG success on single-hop queries.}

\begin{itemize}
    \item \textbf{Query:} For the contingency plan development task, retrieve the outputs produced by the resource allocation strategy implementation.

    \item \textbf{Ground truth:} \texttt{tiered resource pools}.

    \item \textbf{Retrieved context:} The retrieved context contains the key triple
    \path{contingency_plan_development_resource_allocation_strategy_implementation}\texttt{ produces\_output }\path{contingency_plan_development_tiered_resource_pools},along with other resource-allocation-related triples.

    \item \textbf{Model output:} The model correctly returns
    \path{contingency_plan_development_tiered_resource_pools}.

    \item \textbf{Analysis:} The required evidence appears as a directly retrieved one-hop triple. This suggests that GraphRAG can handle simple local triple matching, but may struggle with multi-hop traversal or global graph structure.
\end{itemize}

\paragraph{ToG-RAG success with direct relational evidence.}

\begin{itemize}
    \item \textbf{Query:} How many entities, and what are their IDs, are such that IRS Clearing Members shall only reduce a call for performance bond?

    \item \textbf{Ground truth:} Count: \texttt{1}; entity: \texttt{through the receipt of performance bond deposits}.

    \item \textbf{Retrieved context:} The retrieved context contains a relevant document passage about performance bond deposits and the key triple
    \texttt{(IRS Clearing Members, shall only reduce a call for performance bond, through the receipt of performance bond deposits)}.

    \item \textbf{Model output:} The model identifies the retrieved relation and correctly answers \texttt{1}.

    \item \textbf{Analysis:} This case succeeds because the exact relational evidence is retrieved. It does not require exhaustive traversal or aggregation over a large result set.
\end{itemize}

\paragraph{ToG-RAG failure on aggregation under incomplete retrieval.}

\begin{itemize}
    \item \textbf{Query:} How many tools does the Hawaii last minute getaway task use?

    \item \textbf{Ground truth:} \texttt{12}.

    \item \textbf{Retrieved context:} ToG-RAG retrieves several relevant triples, such as
    \path{hawaii_last_minute_getaway_task}
\texttt{ uses\_tool }
\path{hawaii_last_minute_getaway_check_amenities}
and
\path{hawaii_last_minute_getaway_task}
\texttt{ uses\_tool }
\path{hawaii_last_minute_getaway_search_hotels},
    together with a long dialogue context about hotel amenities and travel planning.

    \item \textbf{Model output:} The model counts only the retrieved tools and returns \texttt{3}.

    \item \textbf{Analysis:} This is a typical aggregation failure caused by retrieval truncation. Exact counting requires the complete set of \texttt{uses\_tool} edges, but top-$k$ retrieval only exposes partial evidence.
\end{itemize}

\paragraph{HippoRAG2 lucky hit without relevant retrieval.}

\begin{itemize}
    \item \textbf{Query:} For the visual cube representation task, how many outcomes does the draw front face step achieve?

    \item \textbf{Ground truth:} \texttt{1}.

    \item \textbf{Retrieved context:} The retrieved contexts are unrelated to the target task, including CrewAI documentation and shadcn/ui code snippets. No relevant knowledge graph evidence about the visual cube representation task is retrieved.

    \item \textbf{Model output:} The model answers that the draw front face step produces exactly one outcome, namely a square serving as the front face of the cube, and returns \texttt{1}.

    \item \textbf{Analysis:} Although the answer matches the ground truth, it is not grounded in the retrieved evidence. The model likely relies on commonsense reasoning about drawing a cube, so this should be interpreted as a lucky hit rather than successful graph reasoning.
\end{itemize}

\paragraph{HippoRAG2 failure on exact two-hop structural queries.}

\begin{itemize}
    \item \textbf{Query:} Find the entity with ID \texttt{mineral\_information\_server} and the entities connected to it by exactly two \texttt{uses\_server} hops, and return their IDs.

    \item \textbf{Ground truth:} \texttt{geological information server} and \texttt{time scale server}.

    \item \textbf{Retrieved context:} HippoRAG2 retrieves unrelated conversational logs and server troubleshooting content, including WildFly connection errors and other MCP-related dialogue. The retrieved context does not mention \texttt{mineral\_information\_server} or the required \texttt{uses\_server} paths.

    \item \textbf{Model output:} The model states that no relevant entity or relation can be found and returns \texttt{None}.

    \item \textbf{Analysis:} This case highlights the weakness of text-first retrieval pipelines on precise graph queries that require exact entity matching and path preservation.
\end{itemize}

%% file: acl_latex.bib
@inproceedings{YinWFS23,
  title={EFO k-cqa: Towards knowledge graph complex query answering beyond set operation},
  author={Yin, Hang and Wang, Zihao and Fei, Weizhi and Song, Yangqiu},
  booktitle={Proceedings of the 31st ACM SIGKDD Conference on Knowledge Discovery and Data Mining V. 2},
  pages={5876--5887},
  year={2025}
}

@inproceedings{WangYS21,
  title={Benchmarking the Combinatorial Generalizability of Complex Query Answering on Knowledge Graphs},
  author={Wang, Zihao and Yin, Hang and Song, Yangqiu},
  booktitle={Thirty-fifth Conference on Neural Information Processing Systems Datasets and Benchmarks Track (Round 2)}
}

@inproceedings{BaiLLYYS23,
  author       = {Jiaxin Bai and
                  Chen Luo and
                  Zheng Li and
                  Qingyu Yin and
                  Bing Yin and
                  Yangqiu Song},
  editor       = {Ambuj K. Singh and
                  Yizhou Sun and
                  Leman Akoglu and
                  Dimitrios Gunopulos and
                  Xifeng Yan and
                  Ravi Kumar and
                  Fatma Ozcan and
                  Jieping Ye},
  title        = {Knowledge Graph Reasoning over Entities and Numerical Values},
  booktitle    = {Proceedings of the 29th {ACM} {SIGKDD} Conference on Knowledge Discovery
                  and Data Mining, {KDD} 2023, Long Beach, CA, USA, August 6-10, 2023},
  pages        = {57--68},
  publisher    = {{ACM}},
  year         = {2023},
  url          = {https://doi.org/10.1145/3580305.3599399},
  doi          = {10.1145/3580305.3599399},
  bibsource    = {dblp computer science bibliography, https://dblp.org}
}

@inproceedings{DemirWLNH23,
  author       = {Caglar Demir and
                  Michel Wiebesiek and
                  Renzhong Lu and
                  Axel{-}Cyrille Ngonga Ngomo and
                  Stefan Heindorf},
  editor       = {Danai Koutra and
                  Claudia Plant and
                  Manuel Gomez Rodriguez and
                  Elena Baralis and
                  Francesco Bonchi},
  title        = {LitCQD: Multi-hop Reasoning in Incomplete Knowledge Graphs with Numeric
                  Literals},
  booktitle    = {Machine Learning and Knowledge Discovery in Databases: Research Track
                  - European Conference, {ECML} {PKDD} 2023, Turin, Italy, September
                  18-22, 2023, Proceedings, Part {III}},
  series       = {Lecture Notes in Computer Science},
  volume       = {14171},
  pages        = {617--633},
  publisher    = {Springer},
  year         = {2023},
  url          = {https://doi.org/10.1007/978-3-031-43418-1\_37},
  doi          = {10.1007/978-3-031-43418-1\_37},
  bibsource    = {dblp computer science bibliography, https://dblp.org}
}

@inproceedings{Bai0LYS24,
  author       = {Jiaxin Bai and
                  Chen Luo and
                  Zheng Li and
                  Qingyu Yin and
                  Yangqiu Song},
  editor       = {Ricardo Baeza{-}Yates and
                  Francesco Bonchi},
  title        = {Understanding Inter-Session Intentions via Complex Logical Reasoning},
  booktitle    = {Proceedings of the 30th {ACM} {SIGKDD} Conference on Knowledge Discovery
                  and Data Mining, {KDD} 2024, Barcelona, Spain, August 25-29, 2024},
  pages        = {71--82},
  publisher    = {{ACM}},
  year         = {2024},
  url          = {https://doi.org/10.1145/3637528.3671808},
  doi          = {10.1145/3637528.3671808},
  bibsource    = {dblp computer science bibliography, https://dblp.org}
}

@misc{gao2025unifying,
      title={Unifying Deductive and Abductive Reasoning in Knowledge Graphs with Masked Diffusion Model}, 
      author={Yisen Gao and Jiaxin Bai and Yi Huang and Xingcheng Fu and Qingyun Sun and Yangqiu Song},
      year={2025},
      eprint={2510.11462},
      archivePrefix={arXiv},
      primaryClass={cs.AI},
      url={https://arxiv.org/abs/2510.11462}, 
}

@article{yisen2025controllable,
  author       = {Yisen Gao and
                  Jiaxin Bai and
                  Tianshi Zheng and
                  Qingyun Sun and
                  Ziwei Zhang and
                  Jianxin Li and
                  Yangqiu Song and
                  Xingcheng Fu},
  title        = {Controllable Logical Hypothesis Generation for Abductive Reasoning
                  in Knowledge Graphs},
  journal      = {CoRR},
  volume       = {abs/2505.20948},
  year         = {2025},
  url          = {https://doi.org/10.48550/arXiv.2505.20948},
  doi          = {10.48550/ARXIV.2505.20948},
  eprinttype    = {arXiv},
  eprint       = {2505.20948},
  bibsource    = {dblp computer science bibliography, https://dblp.org}
}

@inproceedings{Zhu0Z022,
  author       = {Zhaocheng Zhu and
                  Mikhail Galkin and
                  Zuobai Zhang and
                  Jian Tang},
  editor       = {Kamalika Chaudhuri and
                  Stefanie Jegelka and
                  Le Song and
                  Csaba Szepesv{\'{a}}ri and
                  Gang Niu and
                  Sivan Sabato},
  title        = {Neural-Symbolic Models for Logical Queries on Knowledge Graphs},
  booktitle    = {International Conference on Machine Learning, {ICML} 2022, 17-23 July
                  2022, Baltimore, Maryland, {USA}},
  series       = {Proceedings of Machine Learning Research},
  volume       = {162},
  pages        = {27454--27478},
  publisher    = {{PMLR}},
  year         = {2022},
  url          = {https://proceedings.mlr.press/v162/zhu22c.html},
  bibsource    = {dblp computer science bibliography, https://dblp.org}
}

@inproceedings{Mikhail0001DWH22,
  author       = {Mikhail Galkin and
                  Etienne G. Denis and
                  Jiapeng Wu and
                  William L. Hamilton},
  title        = {NodePiece: Compositional and Parameter-Efficient Representations of
                  Large Knowledge Graphs},
  booktitle    = {The Tenth International Conference on Learning Representations, {ICLR}
                  2022, Virtual Event, April 25-29, 2022},
  publisher    = {OpenReview.net},
  year         = {2022},
  url          = {https://openreview.net/forum?id=xMJWUKJnFSw},
  bibsource    = {dblp computer science bibliography, https://dblp.org}
}

@inproceedings{ChenHS22,
  author       = {Xuelu Chen and
                  Ziniu Hu and
                  Yizhou Sun},
  title        = {Fuzzy Logic Based Logical Query Answering on Knowledge Graphs},
  booktitle    = {Thirty-Sixth {AAAI} Conference on Artificial Intelligence, {AAAI}
                  2022, Thirty-Fourth Conference on Innovative Applications of Artificial
                  Intelligence, {IAAI} 2022, The Twelveth Symposium on Educational Advances
                  in Artificial Intelligence, {EAAI} 2022 Virtual Event, February 22
                  - March 1, 2022},
  pages        = {3939--3948},
  publisher    = {{AAAI} Press},
  year         = {2022},
  url          = {https://doi.org/10.1609/aaai.v36i4.20310},
  doi          = {10.1609/AAAI.V36I4.20310},
  bibsource    = {dblp computer science bibliography, https://dblp.org}
}

@inproceedings{ArakelyanDMC21,
  author       = {Erik Arakelyan and
                  Daniel Daza and
                  Pasquale Minervini and
                  Michael Cochez},
  title        = {Complex Query Answering with Neural Link Predictors},
  booktitle    = {9th International Conference on Learning Representations, {ICLR} 2021,
                  Virtual Event, Austria, May 3-7, 2021},
  publisher    = {OpenReview.net},
  year         = {2021},
  url          = {https://openreview.net/forum?id=Mos9F9kDwkz},
  bibsource    = {dblp computer science bibliography, https://dblp.org}
}

@inproceedings{ZhangWCJW21,
  author       = {Zhanqiu Zhang and
                  Jie Wang and
                  Jiajun Chen and
                  Shuiwang Ji and
                  Feng Wu},
  editor       = {Marc'Aurelio Ranzato and
                  Alina Beygelzimer and
                  Yann N. Dauphin and
                  Percy Liang and
                  Jennifer Wortman Vaughan},
  title        = {ConE: Cone Embeddings for Multi-Hop Reasoning over Knowledge Graphs},
  booktitle    = {Advances in Neural Information Processing Systems 34: Annual Conference
                  on Neural Information Processing Systems 2021, NeurIPS 2021, December
                  6-14, 2021, virtual},
  pages        = {19172--19183},
  year         = {2021},
  url          = {https://proceedings.neurips.cc/paper/2021/hash/a0160709701140704575d499c997b6ca-Abstract.html},
  bibsource    = {dblp computer science bibliography, https://dblp.org}
}

@inproceedings{RenHL20,
  author       = {Hongyu Ren and
                  Weihua Hu and
                  Jure Leskovec},
  title        = {Query2box: Reasoning over Knowledge Graphs in Vector Space Using Box
                  Embeddings},
  booktitle    = {8th International Conference on Learning Representations, {ICLR} 2020,
                  Addis Ababa, Ethiopia, April 26-30, 2020},
  publisher    = {OpenReview.net},
  year         = {2020},
  url          = {https://openreview.net/forum?id=BJgr4kSFDS},
  bibsource    = {dblp computer science bibliography, https://dblp.org}
}

@inproceedings{HamiltonBZJL18,
  author       = {William L. Hamilton and
                  Payal Bajaj and
                  Marinka Zitnik and
                  Dan Jurafsky and
                  Jure Leskovec},
  editor       = {Samy Bengio and
                  Hanna M. Wallach and
                  Hugo Larochelle and
                  Kristen Grauman and
                  Nicol{\`{o}} Cesa{-}Bianchi and
                  Roman Garnett},
  title        = {Embedding Logical Queries on Knowledge Graphs},
  booktitle    = {Advances in Neural Information Processing Systems 31: Annual Conference
                  on Neural Information Processing Systems 2018, NeurIPS 2018, December
                  3-8, 2018, Montr{\'{e}}al, Canada},
  pages        = {2030--2041},
  year         = {2018},
  url          = {https://proceedings.neurips.cc/paper/2018/hash/ef50c335cca9f340bde656363ebd02fd-Abstract.html},
  bibsource    = {dblp computer science bibliography, https://dblp.org}
}

@inproceedings{hu2024privacy,
author = {Hu, Qi and Li, Haoran and Bai, Jiaxin and Wang, Zihao and Song, Yangqiu},
title = {Privacy-Preserved Neural Graph Databases},
year = {2024},
isbn = {9798400704901},
publisher = {Association for Computing Machinery},
address = {New York, NY, USA},
url = {https://doi.org/10.1145/3637528.3671678},
doi = {10.1145/3637528.3671678},
booktitle = {Proceedings of the 30th ACM SIGKDD Conference on Knowledge Discovery and Data Mining},
pages = {1108–1118},
numpages = {11},
location = {Barcelona, Spain},
series = {KDD '24}
}

@article{baitop,
  author       = {Jiaxin Bai and
                  Zihao Wang and
                  Yukun Zhou and
                  Hang Yin and
                  Weizhi Fei and
                  Qi Hu and
                  Zheye Deng and
                  Jiayang Cheng and
                  Tianshi Zheng and
                  Hong Ting Tsang and
                  Yisen Gao and
                  Zhongwei Xie and
                  Yufei Li and
                  Lixin Fan and
                  Binhang Yuan and
                  Wei Wang and
                  Lei Chen and
                  Xiaofang Zhou and
                  Yangqiu Song},
  title        = {Top Ten Challenges Towards Agentic Neural Graph Databases},
  journal      = {{IEEE} Data Eng. Bull.},
  volume       = {49},
  number       = {1},
  pages        = {104--123},
  year         = {2025},
  url          = {http://sites.computer.org/debull/A25mar/p104.pdf},
  bibsource    = {dblp computer science bibliography, https://dblp.org}
}

@InProceedings{pmlr-v198-besta22a,
  title = 	 {Neural Graph Databases},
  author =       {Besta, Maciej and Iff, Patrick and Scheidl, Florian and Osawa, Kazuki and Dryden, Nikoli and Podstawski, Michal and Chen, Tiancheng and Hoefler, Torsten},
  booktitle = 	 {Proceedings of the First Learning on Graphs Conference},
  pages = 	 {31:1--31:38},
  year = 	 {2022},
  editor = 	 {Rieck, Bastian and Pascanu, Razvan},
  volume = 	 {198},
  series = 	 {Proceedings of Machine Learning Research},
  month = 	 {09--12 Dec},
  publisher =    {PMLR},
  url = 	 {https://proceedings.mlr.press/v198/besta22a.html}
}

@misc{ren2023neuralgraphreasoningcomplex,
      title={Neural Graph Reasoning: Complex Logical Query Answering Meets Graph Databases}, 
      author={Hongyu Ren and Mikhail Galkin and Michael Cochez and Zhaocheng Zhu and Jure Leskovec},
      year={2023},
      eprint={2303.14617},
      archivePrefix={arXiv},
      primaryClass={cs.DB},
      url={https://arxiv.org/abs/2303.14617}, 
}

@article{10.14778/3574245.3574270,
author = {Sz\'{a}rnyas, G\'{a}bor and Waudby, Jack and Steer, Benjamin A. and Szak\'{a}llas, D\'{a}vid and Birler, Altan and Wu, Mingxi and Zhang, Yuchen and Boncz, Peter},
title = {The LDBC Social Network Benchmark: Business Intelligence Workload},
year = {2022},
issue_date = {December 2022},
publisher = {VLDB Endowment},
volume = {16},
number = {4},
issn = {2150-8097},
url = {https://doi.org/10.14778/3574245.3574270},
doi = {10.14778/3574245.3574270},
journal = {Proc. VLDB Endow.},
month = dec,
pages = {877–890},
numpages = {14}
}

@article{10.14778/3746405.3746424,
author = {Qi, Shipeng and Tong, Bing and Hu, Jiatao and Lin, Heng and Pang, Yue and Yuan, Wei and Lyu, Songlin and Guo, Zhihui and Huang, Ke and Ba, Xujin and Yin, Qiang and Shen, Youren and Zhou, Yan and Lv, Tao and Li, Jia and Zou, Lei and Wu, Yongwei and Sz\'{a}rnyas, G\'{a}bor and Zhu, Xiaowei and Chen, Wenguang and Hong, Chuntao},
title = {The LDBC Financial Benchmark: Transaction Workload},
year = {2025},
issue_date = {May 2025},
publisher = {VLDB Endowment},
volume = {18},
number = {9},
issn = {2150-8097},
url = {https://doi.org/10.14778/3746405.3746424},
doi = {10.14778/3746405.3746424},
journal = {Proc. VLDB Endow.},
month = may,
pages = {3007–3020},
numpages = {14}
}

@article{chandak2022building,
  title={Building a knowledge graph to enable precision medicine},
  author={Chandak, Payal and Huang, Kexin and Zitnik, Marinka},
  journal={Nature Scientific Data},
  doi={https://doi.org/10.1038/s41597-023-01960-3},
  URL={https://www.nature.com/articles/s41597-023-01960-3},
  year={2023}
}

@misc{bai2025autoschemakgautonomousknowledgegraph,
      title={AutoSchemaKG: Autonomous Knowledge Graph Construction through Dynamic Schema Induction from Web-Scale Corpora}, 
      author={Jiaxin Bai and Wei Fan and Qi Hu and Qing Zong and Chunyang Li and Hong Ting Tsang and Hongyu Luo and Yauwai Yim and Haoyu Huang and Xiao Zhou and Feng Qin and Tianshi Zheng and Xi Peng and Xin Yao and Huiwen Yang and Leijie Wu and Yi Ji and Gong Zhang and Renhai Chen and Yangqiu Song},
      year={2025},
      eprint={2505.23628},
      archivePrefix={arXiv},
      primaryClass={cs.CL},
      url={https://arxiv.org/abs/2505.23628}, 
}

@misc{xu2025toucan,
      title={TOUCAN: Synthesizing 1.5M Tool-Agentic Data from Real-World MCP Environments}, 
      author={Zhangchen Xu and Adriana Meza Soria and Shawn Tan and Anurag Roy and Ashish Sunil Agrawal and Radha Poovendran and Rameswar Panda},
      year={2025},
      eprint={2510.01179},
      archivePrefix={arXiv},
      primaryClass={cs.LG},
      url={https://arxiv.org/abs/2510.01179}, 
}

@misc{bai2025longbenchv2deeperunderstanding,
      title={LongBench v2: Towards Deeper Understanding and Reasoning on Realistic Long-context Multitasks}, 
      author={Yushi Bai and Shangqing Tu and Jiajie Zhang and Hao Peng and Xiaozhi Wang and Xin Lv and Shulin Cao and Jiazheng Xu and Lei Hou and Yuxiao Dong and Jie Tang and Juanzi Li},
      year={2025},
      eprint={2412.15204},
      archivePrefix={arXiv},
      primaryClass={cs.CL},
      url={https://arxiv.org/abs/2412.15204}, 
}

@inproceedings{10.5555/3692070.3694094,
author = {Villalobos, Pablo and Ho, Anson and Sevilla, Jaime and Besiroglu, Tamay and Heim, Lennart and Hobbhahn, Marius},
title = {Position: will we run out of data? limits of LLM scaling based on human-generated data},
year = {2024},
publisher = {JMLR.org},
booktitle = {Proceedings of the 41st International Conference on Machine Learning},
articleno = {2024},
numpages = {22},
location = {Vienna, Austria},
series = {ICML'24}
}

@misc{reddit_graph_db,
  author       = {{Reddit User}},
  title        = {{Can someone explain a graph database as compared to a relational database?}},
  howpublished = {\url{https://www.reddit.com/r/cscareerquestions/comments/17ru9l3/can_someone_explain_a_graph_database_as_compared/}},
  year         = {2023},
  note         = {Accessed: 2026-02-08},
  publisher    = {Reddit}
}

@misc{neo4j_why_graph,
  author       = {{Neo4j}},
  title        = {{Why Graph Databases?}},
  howpublished = {\url{https://neo4j.com/why-graph-databases/}},
  year         = {n.d.},
  note         = {Accessed: 2026-02-08}
}

@misc{lai2025graphyourdataendtoendmodeling,
      title={Graphy'our Data: Towards End-to-End Modeling, Exploring and Generating Report from Raw Data}, 
      author={Longbin Lai and Changwei Luo and Yunkai Lou and Mingchen Ju and Zhengyi Yang},
      year={2025},
      eprint={2502.16868},
      archivePrefix={arXiv},
      primaryClass={cs.DB},
      url={https://arxiv.org/abs/2502.16868}, 
}

@misc{kim2024cartepretrainingtransfertabular,
      title={CARTE: Pretraining and Transfer for Tabular Learning}, 
      author={Myung Jun Kim and Léo Grinsztajn and Gaël Varoquaux},
      year={2024},
      eprint={2402.16785},
      archivePrefix={arXiv},
      primaryClass={cs.LG},
      url={https://arxiv.org/abs/2402.16785}, 
}

@misc{deepseekai2025deepseekv32pushingfrontieropen,
      title={DeepSeek-V3.2: Pushing the Frontier of Open Large Language Models}, 
      author={DeepSeek-AI and Aixin Liu and Aoxue Mei and Bangcai Lin and Bing Xue and Bingxuan Wang and Bingzheng Xu and Bochao Wu and Bowei Zhang and Chaofan Lin and Chen Dong and Chengda Lu and Chenggang Zhao and Chengqi Deng and Chenhao Xu and Chong Ruan and Damai Dai and Daya Guo and Dejian Yang and Deli Chen and Erhang Li and Fangqi Zhou and Fangyun Lin and Fucong Dai and Guangbo Hao and Guanting Chen and Guowei Li and H. Zhang and Hanwei Xu and Hao Li and Haofen Liang and Haoran Wei and Haowei Zhang and Haowen Luo and Haozhe Ji and Honghui Ding and Hongxuan Tang and Huanqi Cao and Huazuo Gao and Hui Qu and Hui Zeng and Jialiang Huang and Jiashi Li and Jiaxin Xu and Jiewen Hu and Jingchang Chen and Jingting Xiang and Jingyang Yuan and Jingyuan Cheng and Jinhua Zhu and Jun Ran and Junguang Jiang and Junjie Qiu and Junlong Li and Junxiao Song and Kai Dong and Kaige Gao and Kang Guan and Kexin Huang and Kexing Zhou and Kezhao Huang and Kuai Yu and Lean Wang and Lecong Zhang and Lei Wang and Liang Zhao and Liangsheng Yin and Lihua Guo and Lingxiao Luo and Linwang Ma and Litong Wang and Liyue Zhang and M. S. Di and M. Y Xu and Mingchuan Zhang and Minghua Zhang and Minghui Tang and Mingxu Zhou and Panpan Huang and Peixin Cong and Peiyi Wang and Qiancheng Wang and Qihao Zhu and Qingyang Li and Qinyu Chen and Qiushi Du and Ruiling Xu and Ruiqi Ge and Ruisong Zhang and Ruizhe Pan and Runji Wang and Runqiu Yin and Runxin Xu and Ruomeng Shen and Ruoyu Zhang and S. H. Liu and Shanghao Lu and Shangyan Zhou and Shanhuang Chen and Shaofei Cai and Shaoyuan Chen and Shengding Hu and Shengyu Liu and Shiqiang Hu and Shirong Ma and Shiyu Wang and Shuiping Yu and Shunfeng Zhou and Shuting Pan and Songyang Zhou and Tao Ni and Tao Yun and Tian Pei and Tian Ye and Tianyuan Yue and Wangding Zeng and Wen Liu and Wenfeng Liang and Wenjie Pang and Wenjing Luo and Wenjun Gao and Wentao Zhang and Xi Gao and Xiangwen Wang and Xiao Bi and Xiaodong Liu and Xiaohan Wang and Xiaokang Chen and Xiaokang Zhang and Xiaotao Nie and Xin Cheng and Xin Liu and Xin Xie and Xingchao Liu and Xingkai Yu and Xingyou Li and Xinyu Yang and Xinyuan Li and Xu Chen and Xuecheng Su and Xuehai Pan and Xuheng Lin and Xuwei Fu and Y. Q. Wang and Yang Zhang and Yanhong Xu and Yanru Ma and Yao Li and Yao Li and Yao Zhao and Yaofeng Sun and Yaohui Wang and Yi Qian and Yi Yu and Yichao Zhang and Yifan Ding and Yifan Shi and Yiliang Xiong and Ying He and Ying Zhou and Yinmin Zhong and Yishi Piao and Yisong Wang and Yixiao Chen and Yixuan Tan and Yixuan Wei and Yiyang Ma and Yiyuan Liu and Yonglun Yang and Yongqiang Guo and Yongtong Wu and Yu Wu and Yuan Cheng and Yuan Ou and Yuanfan Xu and Yuduan Wang and Yue Gong and Yuhan Wu and Yuheng Zou and Yukun Li and Yunfan Xiong and Yuxiang Luo and Yuxiang You and Yuxuan Liu and Yuyang Zhou and Z. F. Wu and Z. Z. Ren and Zehua Zhao and Zehui Ren and Zhangli Sha and Zhe Fu and Zhean Xu and Zhenda Xie and Zhengyan Zhang and Zhewen Hao and Zhibin Gou and Zhicheng Ma and Zhigang Yan and Zhihong Shao and Zhixian Huang and Zhiyu Wu and Zhuoshu Li and Zhuping Zhang and Zian Xu and Zihao Wang and Zihui Gu and Zijia Zhu and Zilin Li and Zipeng Zhang and Ziwei Xie and Ziyi Gao and Zizheng Pan and Zongqing Yao and Bei Feng and Hui Li and J. L. Cai and Jiaqi Ni and Lei Xu and Meng Li and Ning Tian and R. J. Chen and R. L. Jin and S. S. Li and Shuang Zhou and Tianyu Sun and X. Q. Li and Xiangyue Jin and Xiaojin Shen and Xiaosha Chen and Xinnan Song and Xinyi Zhou and Y. X. Zhu and Yanping Huang and Yaohui Li and Yi Zheng and Yuchen Zhu and Yunxian Ma and Zhen Huang and Zhipeng Xu and Zhongyu Zhang and Dongjie Ji and Jian Liang and Jianzhong Guo and Jin Chen and Leyi Xia and Miaojun Wang and Mingming Li and Peng Zhang and Ruyi Chen and Shangmian Sun and Shaoqing Wu and Shengfeng Ye and T. Wang and W. L. Xiao and Wei An and Xianzu Wang and Xiaowen Sun and Xiaoxiang Wang and Ying Tang and Yukun Zha and Zekai Zhang and Zhe Ju and Zhen Zhang and Zihua Qu},
      year={2025},
      eprint={2512.02556},
      archivePrefix={arXiv},
      primaryClass={cs.CL},
      url={https://arxiv.org/abs/2512.02556}, 
}

@misc{yang2025qwen3technicalreport,
      title={Qwen3 Technical Report}, 
      author={An Yang and Anfeng Li and Baosong Yang and Beichen Zhang and Binyuan Hui and Bo Zheng and Bowen Yu and Chang Gao and Chengen Huang and Chenxu Lv and Chujie Zheng and Dayiheng Liu and Fan Zhou and Fei Huang and Feng Hu and Hao Ge and Haoran Wei and Huan Lin and Jialong Tang and Jian Yang and Jianhong Tu and Jianwei Zhang and Jianxin Yang and Jiaxi Yang and Jing Zhou and Jingren Zhou and Junyang Lin and Kai Dang and Keqin Bao and Kexin Yang and Le Yu and Lianghao Deng and Mei Li and Mingfeng Xue and Mingze Li and Pei Zhang and Peng Wang and Qin Zhu and Rui Men and Ruize Gao and Shixuan Liu and Shuang Luo and Tianhao Li and Tianyi Tang and Wenbiao Yin and Xingzhang Ren and Xinyu Wang and Xinyu Zhang and Xuancheng Ren and Yang Fan and Yang Su and Yichang Zhang and Yinger Zhang and Yu Wan and Yuqiong Liu and Zekun Wang and Zeyu Cui and Zhenru Zhang and Zhipeng Zhou and Zihan Qiu},
      year={2025},
      eprint={2505.09388},
      archivePrefix={arXiv},
      primaryClass={cs.CL},
      url={https://arxiv.org/abs/2505.09388}, 
}

@article{huang2025information,
  title={Is the Information Bottleneck Robust Enough? Towards Label-Noise Resistant Information Bottleneck Learning},
  author={Huang, Yi and Sun, Qingyun and Gao, Yisen and Yuan, Haonan and Fu, Xingcheng and Li, Jianxin},
  journal={arXiv preprint arXiv:2512.10573},
  year={2025}
}

@misc{qwen3technicalreport,
      title={Qwen3 Technical Report}, 
      author={Qwen Team},
      year={2025},
      eprint={2505.09388},
      archivePrefix={arXiv},
      primaryClass={cs.CL},
      url={https://arxiv.org/abs/2505.09388}, 
}

@article{hu2020open,
  title={Open graph benchmark: Datasets for machine learning on graphs},
  author={Hu, Weihua and Fey, Matthias and Zitnik, Marinka and Dong, Yuxiao and Ren, Hongyu and Liu, Bowen and Catasta, Michele and Leskovec, Jure},
  journal={Advances in neural information processing systems},
  volume={33},
  pages={22118--22133},
  year={2020}
}

@inproceedings{drummond2006open,
  title={The open world assumption},
  author={Drummond, Nick and Shearer, Rob},
  booktitle={eSI workshop: the closed world of databases meets the open world of the semantic web},
  volume={15},
  pages={1},
  year={2006}
}

@article{gutierrez2025rag,
  title={From rag to memory: Non-parametric continual learning for large language models},
  author={Guti{\'e}rrez, Bernal Jim{\'e}nez and Shu, Yiheng and Qi, Weijian and Zhou, Sizhe and Su, Yu},
  journal={arXiv preprint arXiv:2502.14802},
  year={2025}
}

@article{trivedi2022musique,
  title={MuSiQue: Multihop Questions via Single-hop Question Composition},
  author={Trivedi, Harsh and Balasubramanian, Niranjan and Khot, Tushar and Sabharwal, Ashish},
  journal={Transactions of the Association for Computational Linguistics},
  volume={10},
  pages={539--554},
  year={2022},
  publisher={MIT Press One Broadway, 12th Floor, Cambridge, Massachusetts 02142, USA~…}
}

@inproceedings{10.5555/3737916.3738772,
author = {Wang, Minjie and Gan, Quan and Wipf, David and Cai, Zhenkun and Li, Ning and Tang, Jianheng and Zhang, Yanlin and Zhang, Zizhao and Mao, Zunyao and Song, Yakun and Wang, Yanbo and Li, Jiahang and Zhang, Han and Yang, Guang and Qin, Xiao and Lei, Chuan and Zhang, Muhan and Zhang, Weinan and Faloutsos, Christos and Zhang, Zheng},
title = {4DBInfer: a 4D benchmarking toolbox for graph-centric predictive modeling on RDBs},
year = {2024},
isbn = {9798331314385},
publisher = {Curran Associates Inc.},
address = {Red Hook, NY, USA},
booktitle = {Proceedings of the 38th International Conference on Neural Information Processing Systems},
articleno = {856},
numpages = {38},
location = {Vancouver, BC, Canada},
series = {NIPS '24}
}

@misc{gu2026relbenchv2largescalebenchmark,
      title={RelBench v2: A Large-Scale Benchmark and Repository for Relational Data}, 
      author={Justin Gu and Rishabh Ranjan and Charilaos Kanatsoulis and Haiming Tang and Martin Jurkovic and Valter Hudovernik and Mark Znidar and Pranshu Chaturvedi and Parth Shroff and Fengyu Li and Jure Leskovec},
      year={2026},
      eprint={2602.12606},
      archivePrefix={arXiv},
      primaryClass={cs.LG},
      url={https://arxiv.org/abs/2602.12606}, 
}

@article{zhou2024webarena,
  author       = {Shuyan Zhou and
                  Frank F. Xu and
                  Hao Zhu and
                  Xuhui Zhou and
                  Robert Lo and
                  Abishek Sridhar and
                  Xianyi Cheng and
                  Tianyue Ou and
                  Yonatan Bisk and
                  Daniel Fried and
                  Uri Alon and
                  Graham Neubig},
  title        = {WebArena: A Realistic Web Environment for Building Autonomous Agents},
  journal      = {CoRR},
  volume       = {abs/2307.13854},
  year         = {2024},
  url          = {https://doi.org/10.48550/arXiv.2307.13854},
  doi          = {10.48550/ARXIV.2307.13854},
  eprinttype   = {arXiv},
  eprint       = {2307.13854}
}

@article{yang2024finrobot,
  author       = {Hongyang Yang and
                  Boyu Zhang and
                  Neng Wang and
                  Cheng Guo and
                  Xiaoli Zhang and
                  Likun Lin and
                  Junlin Wang and
                  Tianyu Zhou and
                  Mao Guan and
                  Runjia Zhang and
                  Christina Dan Wang},
  title        = {FinRobot: An Open-Source AI Agent Platform for Financial Applications using Large Language Models},
  journal      = {CoRR},
  volume       = {abs/2405.14767},
  year         = {2024},
  url          = {https://doi.org/10.48550/arXiv.2405.14767},
  doi          = {10.48550/ARXIV.2405.14767},
  eprinttype   = {arXiv},
  eprint       = {2405.14767}
}

@inproceedings{chen2021finqa,
  author       = {Zhiyu Chen and
                  Wenhu Chen and
                  Charese Smiley and
                  Sameena Shah and
                  Iana Borova and
                  Dylan Langdon and
                  Reema Moussa and
                  Matt Beane and
                  Ting{-}Hao Huang and
                  Bryan Routledge and
                  William Yang Wang},
  title        = {{F}in{QA}: A Dataset of Numerical Reasoning over Financial Data},
  booktitle    = {Proceedings of the 2021 Conference on Empirical Methods in Natural Language Processing},
  pages        = {3697--3711},
  year         = {2021},
  doi          = {10.18653/v1/2021.emnlp-main.300},
  url          = {https://aclanthology.org/2021.emnlp-main.300/}
}

@inproceedings{reddy2024docfinqa,
  author       = {Varshini Reddy and
                  Rik Koncel{-}Kedziorski and
                  Viet Dac Lai and
                  Michael Krumdick and
                  Charles Lovering and
                  Chris Tanner},
  title        = {{D}oc{F}in{QA}: A Long{-}Context Financial Reasoning Dataset},
  booktitle    = {Proceedings of the 62nd Annual Meeting of the Association for Computational Linguistics (Volume 2: Short Papers)},
  pages        = {445--458},
  year         = {2024},
  doi          = {10.18653/v1/2024.acl-short.42},
  url          = {https://aclanthology.org/2024.acl-short.42/}
}

@inproceedings{yang2022rethinking,
  author       = {Haotong Yang and
                  Zhouchen Lin and
                  Muhan Zhang},
  title        = {Rethinking Knowledge Graph Evaluation Under the Open{-}World Assumption},
  booktitle    = {Advances in Neural Information Processing Systems 35},
  year         = {2022},
  url          = {https://proceedings.neurips.cc/paper_files/paper/2022/hash/378226e5df7eded3e401de5c9493143c-Abstract-Conference.html}
}

@inproceedings{shi2018openworld,
  author       = {Baoxu Shi and
                  Tim Weninger},
  title        = {Open{-}World Knowledge Graph Completion},
  booktitle    = {Proceedings of the AAAI Conference on Artificial Intelligence},
  volume       = {32},
  number       = {1},
  year         = {2018},
  doi          = {10.1609/aaai.v32i1.11535},
  url          = {https://aaai.org/papers/11535-open-world-knowledge-graph-completion/}
}

@article{guo2023datacleaning,
  author       = {Manping Guo and
                  Yiming Wang and
                  Qiaoning Yang and
                  Rui Li and
                  Yang Zhao and
                  Chenfei Li and
                  Mingbo Zhu and
                  Yao Cui and
                  Xin Jiang and
                  Song Sheng and
                  Qingna Li and
                  Rui Gao},
  title        = {Normal Workflow and Key Strategies for Data Cleaning Toward Real{-}World Data: Viewpoint},
  journal      = {Interactive Journal of Medical Research},
  volume       = {12},
  pages        = {e44310},
  year         = {2023},
  doi          = {10.2196/44310},
  url          = {https://www.i-jmr.org/2023/1/e44310/}
}

@inproceedings{qian2023rtgnn,
  author       = {Siyi Qian and
                  Haochao Ying and
                  Renjun Hu and
                  Jingbo Zhou and
                  Jintai Chen and
                  Danny Z. Chen and
                  Jian Wu},
  title        = {Robust Training of Graph Neural Networks via Noise Governance},
  booktitle    = {Proceedings of the Sixteenth ACM International Conference on Web Search and Data Mining},
  pages        = {607--615},
  year         = {2023},
  doi          = {10.1145/3539597.3570369},
  url          = {https://doi.org/10.1145/3539597.3570369}
}

@article{bryzgalova2025missing,
  author       = {Svetlana Bryzgalova and
                  Sven Lerner and
                  Martin Lettau and
                  Markus Pelger},
  title        = {Missing Financial Data},
  journal      = {The Review of Financial Studies},
  volume       = {38},
  number       = {3},
  pages        = {803--882},
  year         = {2025},
  doi          = {10.1093/rfs/hhae036},
  url          = {https://academic.oup.com/rfs/article/38/3/803/7703273}
}

@article{althubaiti2016infobias,
  author       = {Alaa Althubaiti},
  title        = {Information Bias in Health Research: Definition, Pitfalls, and Adjustment Methods},
  journal      = {Journal of Multidisciplinary Healthcare},
  volume       = {9},
  pages        = {211--217},
  year         = {2016},
  doi          = {10.2147/JMDH.S104807},
  url          = {https://pubmed.ncbi.nlm.nih.gov/27217764/}
}

@article{senger2010misspellings,
  author       = {Christian Senger and
                  Jens Kaltschmidt and
                  Simon P. W. Schmitt and
                  Markus G. Pruszydlo and
                  Walter E. Haefeli},
  title        = {Misspellings in Drug Information System Queries: Characteristics of Drug Name Spelling Errors and Strategies for Their Prevention},
  journal      = {International Journal of Medical Informatics},
  volume       = {79},
  number       = {12},
  pages        = {832--839},
  year         = {2010},
  doi          = {10.1016/j.ijmedinf.2010.09.005},
  url          = {https://pubmed.ncbi.nlm.nih.gov/20951634/}
}

@article{jarnac2025uncertainty,
  author       = {Lucas Jarnac and
                  Yoan Chabot and
                  Miguel Couceiro},
  title        = {Uncertainty Management in the Construction of Knowledge Graphs: A Survey},
  journal      = {Transactions on Graph Data and Knowledge},
  volume       = {3},
  number       = {1},
  pages        = {3:1--3:48},
  year         = {2025},
  doi          = {10.4230/TGDK.3.1.3},
  url          = {https://doi.org/10.4230/tgdk.3.1.3}
}

@misc{zhao2026wikidata,
  author       = {Shixiong Zhao and
                  Hideaki Takeda},
  title        = {Diagnosing and Mitigating Semantic Inconsistencies in Wikidata's Classification Hierarchy},
  year         = {2026},
  note         = {Preprint},
  url          = {https://jglobal.jst.go.jp/en/detail?JGLOBAL_ID=202502215099604894}
}

@misc{abaskohi2026drbenchrealisticbenchmarkenterprise,
      title={DRBench: A Realistic Benchmark for Enterprise Deep Research}, 
      author={Amirhossein Abaskohi and Tianyi Chen and Miguel Muñoz-Mármol and Curtis Fox and Amrutha Varshini Ramesh and Étienne Marcotte and Xing Han Lù and Nicolas Chapados and Spandana Gella and Peter West and Giuseppe Carenini and Christopher Pal and Alexandre Drouin and Issam H. Laradji},
      year={2026},
      eprint={2510.00172},
      archivePrefix={arXiv},
      primaryClass={cs.CL},
      url={https://arxiv.org/abs/2510.00172}, 
}

@inproceedings{kayali2025mind,
  title={Mind the Data Gap: Bridging LLMs to Enterprise Data Integration},
  author={Kayali, Moe and Wenz, Fabian and Tatbul, Nesime and Demiralp, Cagatay},
  booktitle={Proceedings of the 15th Conference on Innovative Data Systems Research (CIDR)},
  year={2025},
  address={Amsterdam, The Netherlands},
  url={https://vldb.org/cidrdb/papers/2025/p34-kayali.pdf}
}

@inproceedings{yu-etal-2018-spider,
    title = "{S}pider: A Large-Scale Human-Labeled Dataset for Complex and Cross-Domain Semantic Parsing and Text-to-{SQL} Task",
    author = "Yu, Tao  and
      Zhang, Rui  and
      Yang, Kai  and
      Yasunaga, Michihiro  and
      Wang, Dongxu  and
      Li, Zifan  and
      Ma, James  and
      Li, Irene  and
      Yao, Qingning  and
      Roman, Shanelle  and
      Zhang, Zilin  and
      Radev, Dragomir",
    editor = "Riloff, Ellen  and
      Chiang, David  and
      Hockenmaier, Julia  and
      Tsujii, Jun{'}ichi",
    booktitle = "Proceedings of the 2018 Conference on Empirical Methods in Natural Language Processing",
    month = oct # "-" # nov,
    year = "2018",
    address = "Brussels, Belgium",
    publisher = "Association for Computational Linguistics",
    url = "https://aclanthology.org/D18-1425/",
    doi = "10.18653/v1/D18-1425",
    pages = "3911--3921"
}

@misc{ozsoy2024text2cypherbridgingnaturallanguage,
      title={Text2Cypher: Bridging Natural Language and Graph Databases}, 
      author={Makbule Gulcin Ozsoy and Leila Messallem and Jon Besga and Gianandrea Minneci},
      year={2024},
      eprint={2412.10064},
      archivePrefix={arXiv},
      primaryClass={cs.LG},
      url={https://arxiv.org/abs/2412.10064}, 
}
